RWTH Aachen

Department of Computer Science
*Technical Report*

# MontiArc – Architectural Modeling of Interactive Distributed and Cyber-Physical Systems

Arne Haber and Jan Oliver Ringert and Bernhard Rumpe

RWTH Aachen University
Software Engineering

# MontiArc

## Architectural Modeling of Interactive Distributed and Cyber-Physical Systems – Language Reference –

**Technical Report AIB-2012-03**

Arne Haber
Jan Oliver Ringert
Bernhard Rumpe

**Abstract**

This report presents MontiArc, a modeling language for the description of Component & Connector architectures. A component is a unit executing computations and/or storing data. Information flow between components is modeled via unidirectional connectors connecting typed, directed ports of the interfaces of components.

Language features of the ADL MontiArc include hierarchical decomposition of components, subtyping by structural inheritance, component type definitions and reference declarations for reuse, generic component types and configurable components, syntactic sugar for connectors, and controlled implicit creation of connections and subcomponent declarations.

This technical report gives an overview of the MontiArc language and is a reference for the MontiArc grammar intended to enable reuse and extension of MontiArc and MontiArc related tools.

MontiArc is implemented using the DSL framework MontiCore. Available tools include an editor with syntax highlighting and code completion as well as a simulation framework with a Java code generator.

**Acknowledgment**

We thank Thomas Kutz for his work on previous versions of this report and his help in implementing parts of MontiArc as it is presented here. We also want to acknowledge Class Pinkernell for supervising the diploma thesis in which initial concepts of MontiArc have been developed three years ago.

We are further grateful to Marita Breuer, Hans Grönniger, Holger Krahn, Martin Schindler, Steven Völkel, Galina Volkova for their work on MontiCore. Finally, we want to thank Christoph Hermann and Thomas Kurpick for providing the technical infrastructure via the SSELab[1] that hosts the MontiArc simulation code generation service.

[1] `http://www.sselab.de`

# Contents

# Chapter 1

# Introduction

Distributed interactive systems are systems that typically consist of multiple autonomous computation units that together – composed to a network of communicating nodes – achieve a common goal. The application areas of distributed and embedded systems [GKL+07] are vastly growing and so is the size of the individual systems. Typically, interactive systems are composed of smaller systems or components, that are to a great extent independent of each other. Some concrete examples of distributed systems are:

- telecommunication systems,
- distributed business applications, e.g., service oriented architectures and services in the cloud,
- logical function nets that are mapped to control devices in embedded systems, e.g., in the automotive and avionic domain, or
- production lines and control units in automation technology and process engineering.

Components interact with each other by exchanging messages via their well defined interfaces. Typical kinds of communication are continuous streams of values produced by sensors, complex data messages, events that are propagated, or simple signals passed between components. Modeling the architecture and distribution of such systems allows early analysis of certain properties like the absence of deadlocks, interface compatibility of connected components, and simulation of effect propagation.

Many systems currently emerge in the new domain of Cyber-Physical Systems [GRSS12]. Cyber-Physical Systems are inherently distributed interacting in various ways using signals, messages and data. However, model based development of Cyber-Physical Systems become particularly interesting, when modelling the context of the software control, i.e. electric and hydraulic signals as well as physical material (streams of fluids or gadgets) to simulate the system under development early. We provide an infrastructure to model these kinds of streams as well.

For the task of modeling distributed interactive systems by capturing elements of their logical or physical distribution we have developed MontiArc. MontiArc is a textual language defined using the DSL framework MontiCore [GKR+06, KRV07b, GKR+07, KRV08, GKR+08, KRV10, Kra10, www12b] and comes with an Eclipse integrated editor. To analyze and simulate the designed systems, MontiArc is extended with a simulation framework that can execute behavior implemented in Java and declaratively attached to MontiArc models. Language tooling including the simulator is available at the MontiArc website [www12a].

MontiArc allows modeling of function nets as described in [GHK+07, GHK+08a, GHK+08b, GKPR08, MPF09]. Using a set of feature views enables tracing of requirements to the logical system architecture to conceptually link artifacts throughout the development process. These feature views are later on composed to a logical system architecture that is deployed to concrete hard- and software components [GHK+08b].

MontiArc has also been extended with delta modeling concepts [CHS10] to Δ-MontiArc [HRRS11, HKR+11a, HKR+11b]. It allows a modular definition of architectural variants starting from a core architecture. Features are added or removed by modular delta models that contain operations to add, remove, or modify elements of an architecture. New variants are generated by a sequence of deltas that transform the core architecture to the desired architectural variant.

In this report we will introduce the rationale behind MontiArc briefly in the following sections on a simple example. The language reference describing MontiArc's language features, the current version of MontiArc's grammar, and its context conditions is given in Chapters 2 and 3. MontiArc's semantics is described in Chapter 4.

## 1.1 MontiArc ADL

MontiArc is a framework for modeling and simulation of software architectures. The domain of the architecture description language (*ADL*) MontiArc are so called information-flow architectures which describe the components of a (software) system and their message-based communication. Following [MT00]'s taxonomy for ADLs a component is a unit which executes computations or stores data. It may have arbitrary complexity and size being a subsystem or a single function. A component has an explicitly defined interface via which it communicates with its environment. MontiArc distinguishes between component type definitions and subcomponent declarations. A *component type* defines the component's interface and its internal structure using subcomponent declarations. A *subcomponent declaration* describes the inclusion of a component of a referenced type.

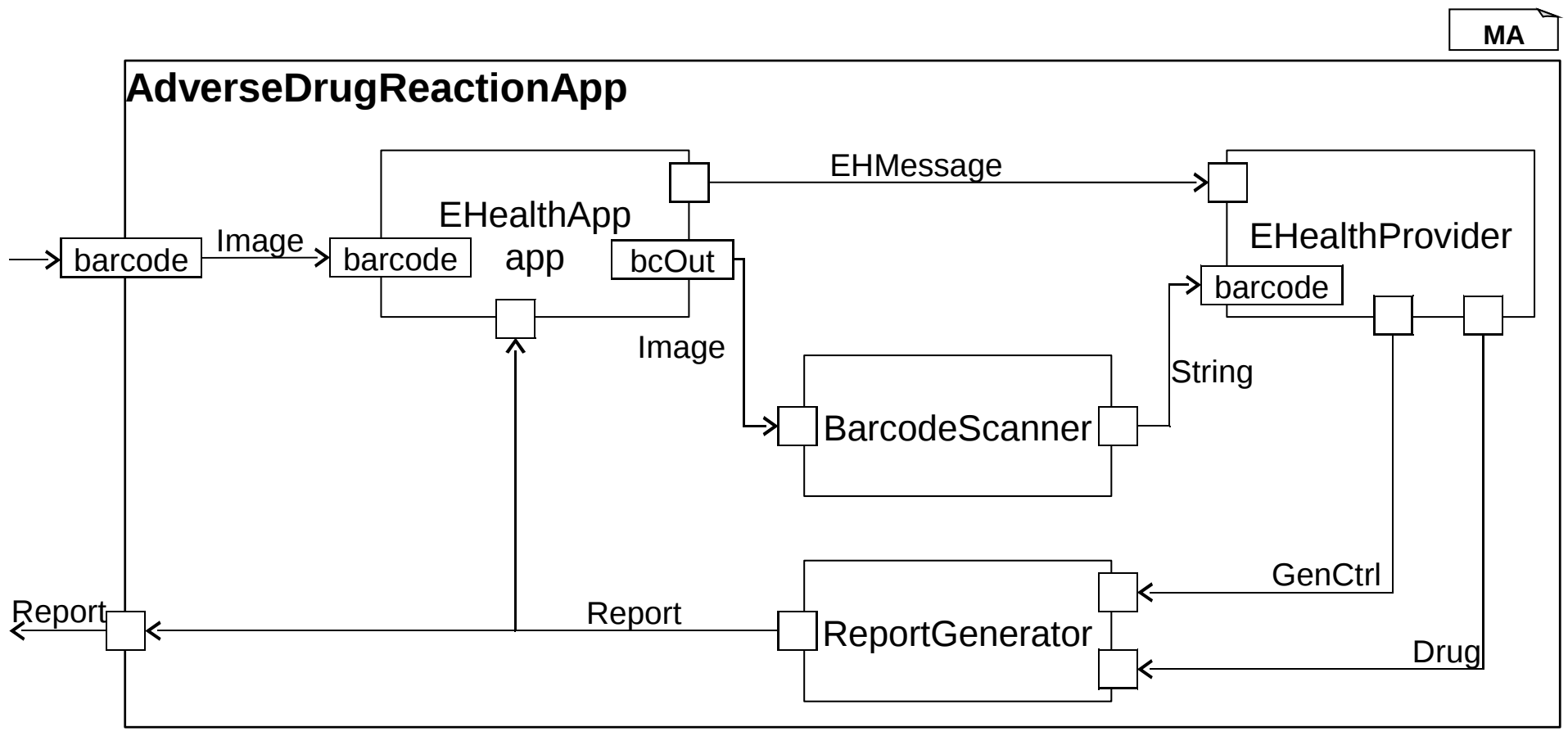

Figure 1.1: Component type `AdverseDrugReactionApp`

As an initial example consider Fig. 1.1 which shows the architecture of a distributed e-health application. The subsystems involved are the application itself receiving the picture of a barcode from the environment, a barcode scanner service extracting the actual barcode, an e-health provider that resolves possible adverse reactions for given medications, and a report generator that produces a report. The illustrated architecture contains all basic elements of the ADL MontiArc. The component `AdverseDrugReactionApp` can only be accessed via messages on the incoming port `barcode` of type `Image`. The computed result will then be sent via an outgoing port of type `Report`. The component `AdverseDrugReactionApp` consists of subcomponents `EHealthApp` named `app`, `BarcodeScanner`, `EHealthProvider` and `ReportGenerator` which are connected to each other and outer ports exposed to the environment via connectors. Connectors are directed and always connect one sending port with an arbitrary number of receiving ports of compatible data types. Two ports can be connected, if either their data types are identical or the data type of the receiver is a supertype of the sender's data type. As shown in the example, naming of ports and subcomponents is optional. Subcomponent `app`, which is an instance of component `EHealthApp`, for example, is explicitly named, while subcomponent `BarcodeScanner` by default gets the name `barcodeScanner` of its referenced component type starting with a lower-case character.

In MontiArc communication is unidirectional. The ADL does not dictate any further communication semantics. Communication can be synchronous or asynchronous based on the system that is modeled. In most cases and in our MontiArc simulation we assume an asynchronous communication which we believe is better suited for modeling parallel and distributed computa-tions. Subcomponent `app` in Fig. 1.1 sends messages to `BarcodeScanner-Service` and `EHealthProvider` whereas indirect responses are received at the incoming port `Report`.

In MontiArc components can either be implemented directly (as *atomic components*) or defined to be a composition of other components. These *decomposed components* are hierarchically structured into further subcomponents (as for example `AdverseDrugReactionApp` in Fig. 1.1) and thus have their behavior derived from the composition of their subcomponents. For atomic components a reaction to incoming messages can be specified directly. Both component kinds can be equally treated in an architecture model, since they both share the same interface definitions and can be treated as black boxes whose observable behavior conforms to their interface. Thus, in Figure 1.1 it is initially non-distinctive, whether the `EHealthProvider` is a structured or atomic component. For different levels of abstractions this can however change during the development of the `AdverseDrugReactionApp` by consecutively refining the structure of its components.

## 1.2 MontiArc Syntax at a Glance

MontiArc is developed as a textual domain specific language (*DSL*) using the MontiCore [GKR+08, GKR+06, KRV07b, KRV07a, KRV08, KRV10, Kra10, www12b] framework. A textually formalized description of the component `AdverseDrugReactionApp` from Fig. 1.1 is given in Lst. 1.2.

Each component is typically defined in its own compilation unit (file). Similar to Java, components are organized in packages (Lst. 1.2, l. 1) which correspond to subfolders in the modelpath. Imports of data types (ll. 4 f) and components (ll. 8–10) refer to other compilation units using the same import mechanism and make their declared names available to be referenced using unqualified names. The component type definition is introduced by the keyword `component` followed by the component's type name (l. 12). Inside the component body, which is delimited by curly brackets, architectural elements may be defined. Among these are interfaces, invariants, subcomponent declarations, definitions of inner component types, and connectors.

The interface of a component, which defines the communication with its environment, is given by ports. Port definitions are introduced by the keyword `port` followed by the communication direction (`in` for incoming, `out` for outgoing) (ll. 17-19). The data type (e.g., `Image`, l. 18) specifies which types of messages can be transmitted via a port. Naming of ports is optional. If only one port of a data type exists on the given level, the port does not need an explicit name, since it then by default gets the name of its type starting with a lower-case character. For example, the port in

MontiArc

```
1 package adra;
2 
3 // import message types
4 import java.awt.Image;
5 import adra.msg.*;
6 
7 // import components
8 import adra.fe.EHealthApp;
9 import adra.be.BarcodeScanner;
10 import adra.be.ReportGenerator;
11 
12 component AdverseDrugReactionApp {
13 
14   autoconnect port;
15   autoinstantiate on;
16 
17   port
18     in Image barcode,
19     out Report;
20 
21   component EHealthProvider {
22     port
23         in EHMessage,
24         in String barcode,
25         out GenCtrl,
26         out Drug;
27   }
28 
29   component EHealthApp app
30     [bcOut -> barcodeScanner.image];
31 
32   component BarcodeScanner;
33 
34   component ReportGenerator;
35 
36   connect barcodeScanner.string -> eHealthProvider.barcode;
37 
38   connect eHealthProvider -> reportGenerator;
39 }
```

Listing 1.2: The component type `AdverseDrugReactionApp` in textual syntax

l. 19 is named `report` like its type , while port `barcode` has an explicit name (l. 18). If the same data type is used for several ports of the same component, these ports have to be explicitly named, because the default name derived from the type would not be unique anymore.

Inner component types are used to create local component type defi-nitions that may be used in the current component exclusively. For example, component `EHealthProvider` is defined as an inner component of the `AdverseDrugReactionApp` (ll. 21–27). The definition of an inner component type starts with the keyword `component` followed by the component's type name. It then has the same structure as the component definition on the top level of the model. To automatically create instances of inner components, auto-instantiation is turned on (l. 15). This way a subcomponent named `eHealthProvider` is created that instantiates the inner component type.

Similarly, a component that is defined in another model can be referenced and instantiated as a subcomponent (ll. 29-34). Like ports, subcomponents have an optional name (e.g., `app`, l. 29) if it should be different from the referenced component type. However, if multiple subcomponents of a certain type exist or if the simple connector form, which will be described below, is to be used, subcomponents have to be named explicitly.

The communication connections between subcomponents' ports and incoming and outgoing ports of the component definition are realized by connectors. For this purpose, MontiArc provides several alternatives that may be combined with each other:

1. Standard connectors are created by the keyword `connect` and connect a single source with an arbitrary number of targets (l. 36). Sources and targets are qualified by the subcomponent name the port is attached to. If more then one target is given they have to be separated by commas.

2. To immediately connect outputs of a subcomponent, simple connectors are placed directly behind the instance name (l. 30). Here, the source `bcOut` is an outgoing port of `EHealthApp`. Other than in standard connector definitions, the left side of a simple connector directly refers to an outgoing port of the referenced component (and is therefore unqualified).

3. The automatic connecting of not yet connected ports with the **same name** and the same unambiguous type is triggered by the keyword phrase `autoconnect port` (l. 14). However, it is not always possible to establish the whole communication graph with this statement, hence connect statements allow to explicitly define connections.

4. To automatically connect all not yet connected ports with the same unambiguous type **disregarding port names**, the keyword phrase `autoconnect type` is used. Depending on the current component definition, its interface, and its subcomponents, the set of ports matched by this phrase differs from the set matched by autoconnect port.

5. Instead of referencing ports of subcomponents as source and target of a (simple) connector, a name of a subcomponent contained in the current component definition may be used as well. This means that all compatible ports of the referenced subcomponents are connected automatically. The connector shown in line 38 connects all outgoing ports of subcomponent `eHealthProvider` with all compatible ports of subcomponent `reportGenerator`.

Thus, the connector in line 36 connects port `string` of subcomponent `barcodeScanner` with port `barcode` of subcomponent `eHealthProvider`. The simple connector in line 30 connects the port `bcOut` of the subcomponent `app` with the port `image` of the subcomponent `barcodeScanner`. Which form of the explicit connectors is used is a matter of taste. Using simple connectors one can only specify outgoing connections of a subcomponent and not its incoming ones.

## 1.3 Communication of MontiArc Components

The definition of inter component communication and the simulation of MontiArc models is based on Focus [BS01, RR11], a framework for developing and modeling distributed systems. Communication in MontiArc is typically asynchronous and event based. It is realized by unidirectional channels which transport elements of a data type that represent events and messages that are passed between components. A channel contains messages in order of their transmission and is mathematically formalized using a stream of messages, e.g., $\langle m_1, m_2, m_3, \dots \rangle$.

In streams the order of messages determines the order of transmission. Concerning the time lag between the messages, no statement can be derived. In Focus, the progress of time can be simulated by explicitly adding tick messages ($\surd$) that to some extent can be treated as ordinary messages. Every equidistant time slice is delimited by a $\surd$. For messages inside a time slice, only the order of the transmission is fixed.

In Focus, there are three different time paradigms. In *timed streams*, a time slice contains an arbitrary but finite number of messages whereas in *time-synchronous streams* at most one message per time slice is transmitted. Untimed streams contain messages only, this way conclusions about the order of messages can be drawn but not on their timing. For the remainder we focus on timed streams which are well supported by the MontiArc simulation framework.

In order to predict the timed behavior of components, a basic understanding of the time model is necessary. In timed communication the time flow is modeled in such a way that components, which have a $\surd$ on each incoming port, consume these $\surd$s and send one $\surd$ on each outgoing port. This mimics synchronized time progress on all channels, whereas the $\surd$s denote a time slice has passed. Please note that message processing in the MontiArc simulation framework enforces realizability of the simulated components. We achieve this by requiring strong causality [BS01, RR11] for components: each component only reacts to the current input after a delay of at least one $\surd$.[1]

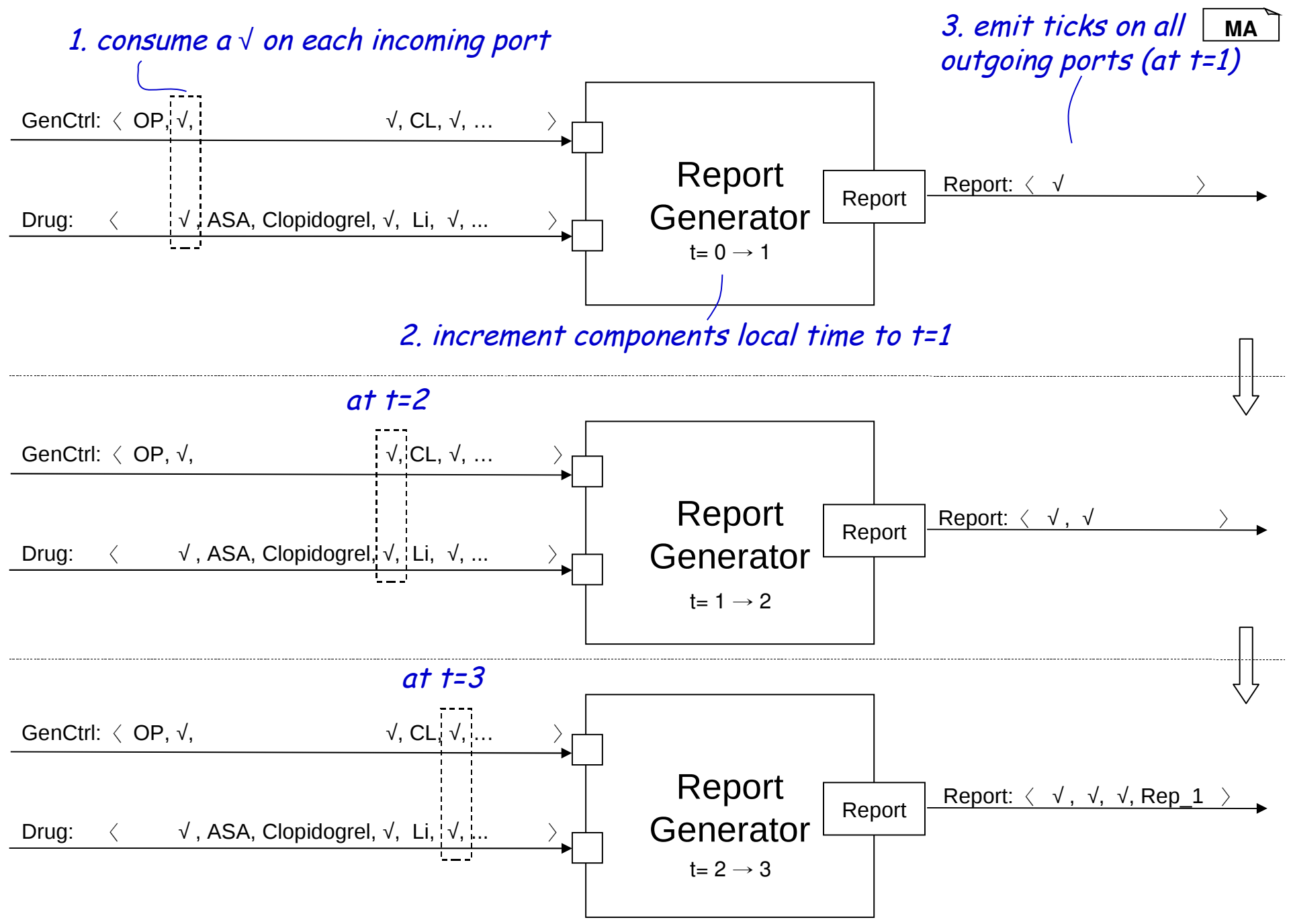

Figure 1.3: Processing of messages in the simulation

Figure 1.3 illustrates how messages are processed by MontiArc components. Messages that are not processable immediately are buffered to not get lost. The message `OP` sent to component `ReportGenerator` is processed immediately inducing the component to open and generate a new report. Processed messages are removed from the message buffer. $\surd$s are only consumed when each incoming port of a component has received an unprocessed $\surd$. This way the report generator has to receive a $\surd$ on port `GenCtrl` and `Drug`. Afterwards the local time of the component is increased by one and a $\surd$ is emitted on each outgoing port. After receiving the messages `ASA` (acetylsalicylic acid) and `Clopidogrel` on port `Drug` in $t = 1$, the component receives message `CL` on port `GenCtrl` and `Li` (lithium) on port `Drug` in $t = 2$. As message `CL` instructs the component to close the report and send it in the next time unit, it emits the generated report in $t = 3$. Hence `Rep_1` contains adverse drug reactions between ASA, Clopidogrel, and Li.

[1]The simulation framework also supports weakly causal components and a delay of at least one $\surd$ in every feedback-loop which also makes a system realizable [RR11].

Please note that components are allowed to access the history of a stream but not the message buffers which would in a timed setting allow (partial) knowledge about the future. This is exactly how weak causality of component interaction is enforced.

The MontiArc simulation framework with a Java code generator, a tutorial, and a set of executable examples is available from [www12a].

# Chapter 2

# MontiArc Language

MontiArc is developed with the DSL framework MontiCore [KRV10] using its language-extension mechanisms to create an expandable architectural description language according to the guidelines presented in [KKP+09]. Its language hierarchy is shown in Fig. 2.1.

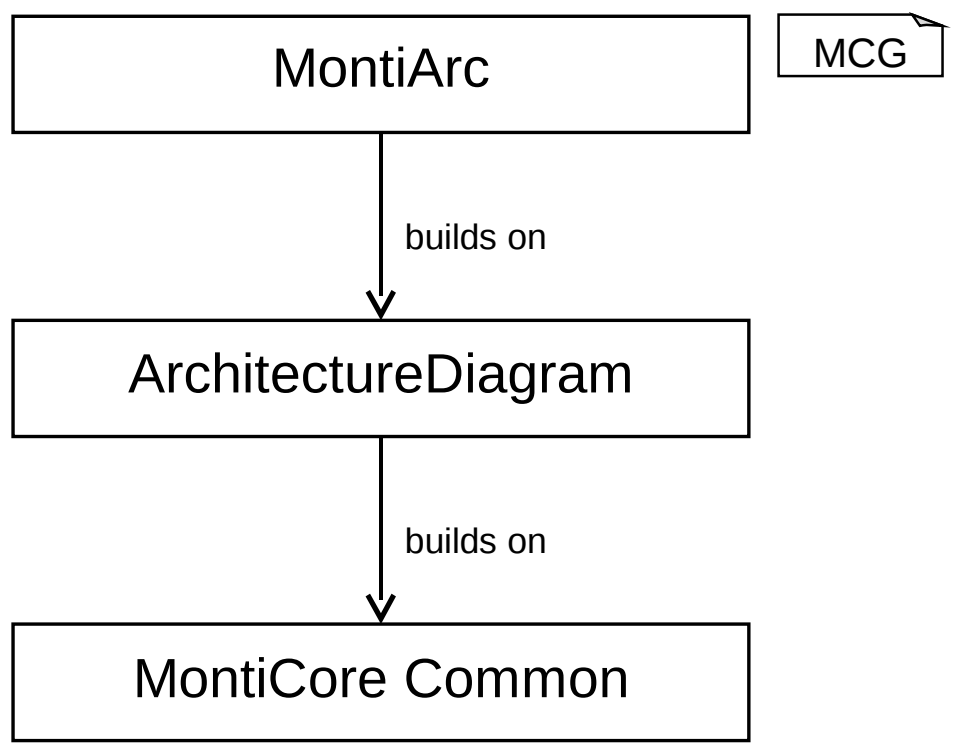

Figure 2.1: MontiCore grammar hierarchy of the MontiArc language.

The `MontiCore Common` grammar serves common modeling language artifacts like stereotypes, cardinalities, and modifiers. Also productions for type references and literals are provided. A detailed description of the used language fundamentals is given in [Sch12], the MontiCore grammar format is described in [GKR+06]. The language `ArchitectureDiagram` (ArcD) serves basic architectural elements and is extended by the `MontiArc` language that is tailored to the event-based simulation of the modeled distributed systems. The main elements and concepts of the two languages are explained in Sections 2.1 and 2.2. Complete grammar definitions are given in the appendix in Lst. B.1 on page 51 and Lst. B.2 on page 57. The MontiCore grammar of both languages will be explained in Sections 2.3 and 2.4.

## 2.1 Architecture Diagram – Basic Elements

The basic elements of the Architecture Diagram language are components with ports, subcomponents, and connectors that unidirectional connect ports of components. A component type definition defines a component type and its interface. The interface of a component is a set of typed and directed ports. The internal structure of components can be defined by referencing subcomponents and composing them via connector definitions.

*Component type definitions* introduce new component types that are identified by unique qualified names and interfaces consisting of sets of ports. A component's ports are either incoming or outgoing ports. Incoming ports allow a component to receive messages of the port's type. In the current implementation MontiArc allows the modeler to use predefined types of the Java language or define new types using UML/P class diagrams [Rum04b, Sch12]. Outgoing ports are a component's mean to communicate with it's environment. These ports are also typed with the type of messages that can be transmitted.

Component type names can be parametrized with type parameters. This concept allows the definition of *generic component types* that use type parameters inside their component's ports definition or as type parameters in references to subcomponents. An example for generic components with type parameters is a delay component that is parametrized with the type of its incoming and delayed outgoing messages.

Besides type parameters MontiArc also supports *configurable components* with configuration parameters. These parameters are values that are set when a component is referenced as a subcomponent or when instantiated as a system inside the MontiArc simulation. An example for a configurable component is a filter component that can be configured with a set of elements that should be filtered out.

*Connectors* define unidirectional communication channels between ports of components. Messages emitted from an outgoing port of one component and forwarded to an incoming port of another component are the only way that components can interact (feedback-loops of a component to itself are also allowed).

MontiArc supports the extension of component definitions by a *structural inheritance* mechanism. A component type definition can extend another component type. The new type inherits all ports, inner component type definitions, and subcomponent references as well as the connectors defined between them.

## 2.2 MontiArc Elements

The language MontiArc is defined as an extension of Architecture Diagrams. It preserves all language concepts and adds:

- invariants on component behavior
- declaration of the component's timing paradigm
- implicit model completion for connections
- implicit reference declaration for subcomponents

*Invariant expressions* can be defined for components and written in almost any external invariant language. With MontiCore's language reuse features [GKR+08] the MontiArc language is currently configured to use Java expressions and OCL expressions as invariants.

The *timing paradigm* of components can be defined as part of the component type definition inside MontiArc. The available paradigms are `timed`, `untimed`, and `timesynchronous` as explained in Sect. 1.3.

MontiArc allows to automatically complete component definitions with connectors according to some predefined rules. The *connectors completion strategies* currently available are `port` to connect ports with matching name and compatible type, `type` to connect ports with matching types, or `off` to disable automatic connection of ports.

In many cases when a component type is defined inside a parent component it is the modelers intention to create a subcomponent by referencing this new component inside the parent component. To *automatically instantiate subcomponents* together with their definition the modeler can enable the `autoinstantiate` concept on the parent's component level.

## 2.3 Architecture Diagram Grammar Walk-Through

We continue by explaining the MontiCore grammar of the architecture diagrams language ArdD. A detailed description of the MontiCore grammar format that is used to define concrete and abstract syntax of a language is found in [GKR+08]. However, the most important concepts of MontiCore grammars, that are needed to understand the given language definitions, are the following. Optional elements are annotated with a question mark `?`, alternatives are separated by `|`, and keywords are given in quotes. If keywords are additionally surrounded by brackets (`[...]`) a Boolean field is created in the abstract syntax that holds true, if this keyword occurs in the concrete syntax. A `*` denotes elements that may occur arbitrary many times.

Grammar for ArcD

```
1 ArcComponent implements ArcElement =
2   Stereotype?
3   "component" Name (instanceName:Name)?
4   ArcComponentHead ArcComponentBody;
```

Listing 2.2: Component type definition production

The root element of an ArcD model is a component type definition. The production `ArcComponent` that is shown in Lst. 2.2 defines the structure of a component type definition. It may be annotated with a stereotype followed by the keyword `component` and a component type name. The optional `instanceName` may be used to create a subcomponent declaration along with the definition of an inner component type. For root component definitions the usage of an instance name is forbidden (c.f. Chapter 3).

Grammar for ArcD

```
1 ArcComponentHead =
2   TypeParameters?
3   ("[" ArcParameter* "]")?
4   ("extends" ReferenceType)?;
```

Listing 2.3: Component head production

Grammar for ArcD

```
1 ArcParameter =
2   Type Name;
```

Listing 2.4: Parameter definition production

The production `ArcComponentHead` is shown in Lst. 2.3. It provides optional definition of `TypeParameters` (c.f. l. 2) that are used to define generic type variables. These variables may serve as port data types in the scope of the component body. A list of `ArcParameters` is enclosed by squared brackets (c.f. l. 3). These are used to define variables with a type and a name that are visible in the scope of the component definition (c.f. Lst. 2.4). The values of these variables are set when a parametrizable component type is used as type of a subcomponent declaration. Finally a component type may extend a super component, its type name is given after the `extends` keyword (c.f. Lst. 2.3 l. 4).

Lst. 2.5 shows the production of a component type body. It contains arbitrary many `ArcElements` that are parenthesized by curly brackets. `ArcElement` is an interface that is implemented by productions that are architectural elements and may occur in a component type definition. Therefore the inner structure of a component type is given by a set of

```
Grammar for ArcD
ArcComponentBody =
  "{"
    ArcElement*
  "}";
```

Listing 2.5: Component body production

ArcElements. To extend this language with more elements that may be part of a component type, new productions that implement this interface may be defined in a subgrammar.

```
Grammar for ArcD
ArcInterface implements ArcElement =
  Stereotype?
  "port" (ArcPort)* ";";
```

Listing 2.6: Interface definition production

```
Grammar for ArcD
ArcPort =
  Stereotype?
  ("in" | "out")
  Type Name?;
```

Listing 2.7: Port definition production

The production ArcInterface that defines the interface definition of a component is given in Lst. 2.6. After an optional stereotype and the keyword port a list of ports is given. A port (c.f. Lst. 2.7) may have a stereotype. After the port's direction, in is used for incoming and out for outgoing ports, the port's data type and an optional name are given.

```
Grammar for ArcD
ArcSubComponent implements ArcElement =
  Stereotype?
  "component"
  ReferenceType
  ("(" ArcConfigurationParameter* ")" )?
  (ArcSubComponentInstance* )? ";";
```

Listing 2.8: Production for subcomponent declarations

The internal structure of decomposed component types is given by subcomponents. The syntax of a subcomponent declaration is defined by the production ArcSubComponent that is shown in Lst. 2.8. After an optional

Grammar for ArcD

```
ArcConfigurationParameter =
  QualifiedName | Literal;
```

Listing 2.9: Configuration parameter production

stereotype and the keyword `component` the type of the subcomponent is given. This is a reference to another component type definition. An optional list of arguments is parenthesized by round brackets. These arguments are used to set configuration parameters of referenced configurable components. As shown in Lst. 2.9 this may be either a reference to an enumeration or a static constant or a variable name (both given by a `QualifiedName`), or a literal value.

Grammar for ArcD

```
ArcSubComponentInstance =
  Name
  ("[" ArcSimpleConnector
  (";" ArcSimpleConnector)* "]")?;
```

Listing 2.10: Production to explicitly name subcomponents with optional simple connectors

Grammar for ArcD

```
ArcSimpleConnector =
  source:QualifiedName "->" targets:QualifiedName
                           ("," targets:QualifiedName)*;
```

Listing 2.11: Simple connector production

To create more then one subcomponent declaration or to assign an explicit name, an optional list of instances is used (c.f. Lst. 2.8 l. 6). The production `ArcSubComponentInstance` is shown in Lst. 2.10. It has a name and an optional colon separated list of simple connectors parenthesized by squared brackets. These simple connectors (c.f. Lst. 2.11) directly connect outgoing ports of the bounded subcomponent declaration with one or more target ports. Please note that `source:QualifiedName` is an extension to normal grammars, where `QualifiedName` is the nonterminal (type) and `source` is the name of the containing variable in the abstract syntax that usually also codes the form of use. Here it allows to distinguish between a source and many targets (see [GKR+06]).

Another way to connect ports of subcomponent declarations or the local interface definition is given by connectors. The syntax is defined by the `ArcConnector` production that is shown in Lst. 2.12. After an optional stereotype and the keyword `connect` the `source` of the connector is given

```
                                    Grammar for ArcD
ArcConnector implements ArcElement=
  Stereotype?
  "connect" source:QualifiedName "->"
  targets:QualifiedName ("," targets:QualifiedName)* ";";
```

Listing 2.12: Connector production

by a qualified name. After an arrow -> one or more comma separated targets are given. Source or target of a connector may be either a port of the current component, a name of a subcomponent declaration, or a port that belongs to a subcomponent declaration. In the last case the port is qualified by the name of the subcomponent to which it belongs.

## 2.4 MontiArc Grammar Walk-Through

```
                                    Grammar for MontiArc
ArcComponentBody =
  "{"
    MontiArcConfig*
    ArcElement*
  "}";
```

Listing 2.13: Component body production in MontiArc

MontiArc extends the ArchitectureDiagram language in two ways. First, it extends the language with configuration elements that have to implement the interface MontiArcConfig. These configuration elements have to be placed before other architectural elements in a component's body. Hence the production ArcComponentBody of the super grammar is overridden as shown in Lst. 2.13.

```
                                    Grammar for MontiArc
MontiArcInvariant implements ArcElement =
  "inv" Name ":" InvariantContent ";";
```

Listing 2.14: Invariant production in MontiArc

Second, to constrain the behavior of a component MontiArc adds invariants defined in OCL/P [Rum04b, Rum04a] or Java to components. This is shown in Lst. 2.14. After the keyword inv the name of the invariant is given followed by its content.

The autoconnect statement defined in production MontiArcAutoConnect (c.f. Lst. 2.15) controls the autoconnect behavior of the component. The following modes are available:

Grammar for MontiArc

```
MontiArcAutoConnect implements MontiArcConfig =
  "autoconnect" Stereotype?
  ("type" | "port" | "off") ";";
```

Listing 2.15: Autoconnect statement in MontiArc

- `type` automatically connect all ports with the same unique type
- `port` automatically connect all ports with the same name and a compatible type
- `off` (default) turns auto connect off

Grammar for MontiArc

```
MontiArcAutoInstantiate implements MontiArcConfig  =
  "autoinstantiate" Stereotype?
  ("on" | "off") ";";
```

Listing 2.16: Autoinstantiate statement in MontiArc

Auto-instantiation is used to automatically create instances of inner component types, if these are not explicitly declared as subcomponents. Lst. 2.16 contains the production defining the syntax of this feature. After the keyword `autoinstantiate` and an optional stereotype the mode is chosen using `on` or `off`, where `off` is the default case.

Grammar for MontiArc

```
MontiArcTimingParadigm implements MontiArcConfig  =
  "behavior" Stereotype?
  ("timed" | "untimed" | "timesynchronous") ";";
```

Listing 2.17: Production to choose a timing paradigm in MontiArc

To denote which timing paradigm (see Sect. 1.3 on page 7) a component implements the production `MontiArcTimingParadigm` is used that is shown in Lst. 2.17. After the keyword `behavior`, and an optional stereotype, the following modes are available:

- `timed` (default) streams are timed and may contain arbitrary many data messages in a time slice
- `untimed` streams are untimed and contain only data messages
- `timesynchronous` streams are timed but contain at most one data message in a time slice, hence all incoming ports of a component are read simultaneously

Each component type may have its own timing paradigm. If subcomponent declarations of a component type definition reference component types with different timing paradigms, the distinct behavior regarding time has to be adapted to achieve a smooth interaction. Introducing up- and down-scaling subcomponent declarations that translate between different timing paradigms in terms of a behavior refinement (see [BS01]) will then serve as adapters between subcomponents with different timing paradigms.

# Chapter 3

# MontiArc Context Conditions

In MontiArc quite a number of context condition checks are implemented in order to verify that a defined MontiArc model is well-formed and to support the modeler with feedback about detected problems. These context conditions are grouped into conditions concerning uniqueness, connections, referential integrity, and conventions. The following sections list these conditions and explain them by means of examples.

## 3.1 Basic Conditions

To define a concept for visibility of identifiers we introduce namespaces to MontiArc that define areas in a model in which names are managed together (c.f. [Kra10, Völ11]). These identifiers are names of ports, subcomponent declarations, generic type variables, configuration parameters, and invariant definitions. In MontiArc we distinguish two different kinds of namespaces. A *component namespace* contains identifiers that are declared within a component type definition. Such a namespace is not hierarchical, hence identifiers defined in a top level namespace are not imported into a contained component namespace. In contrast, an *invariant namespace* that is contained in a component namespace imports all names that are defined within its parent namespace. An invariant namespace may also contain a hierarchical namespace structure according to the language that is used to define the invariant.

An example for namespaces, identifiers, and their visibility is given in Fig. 3.1. The shown component type definition `FilterDelay` contains three namespaces. The top-level namespace belongs to the component type definition itself. It contains the red colored identifiers of the configuration parameter `fil`, the port names `inData` and `delayedAndFiltered`, the subcomponent declaration `f` and `del`, as well as the invariant `isFiltered`.

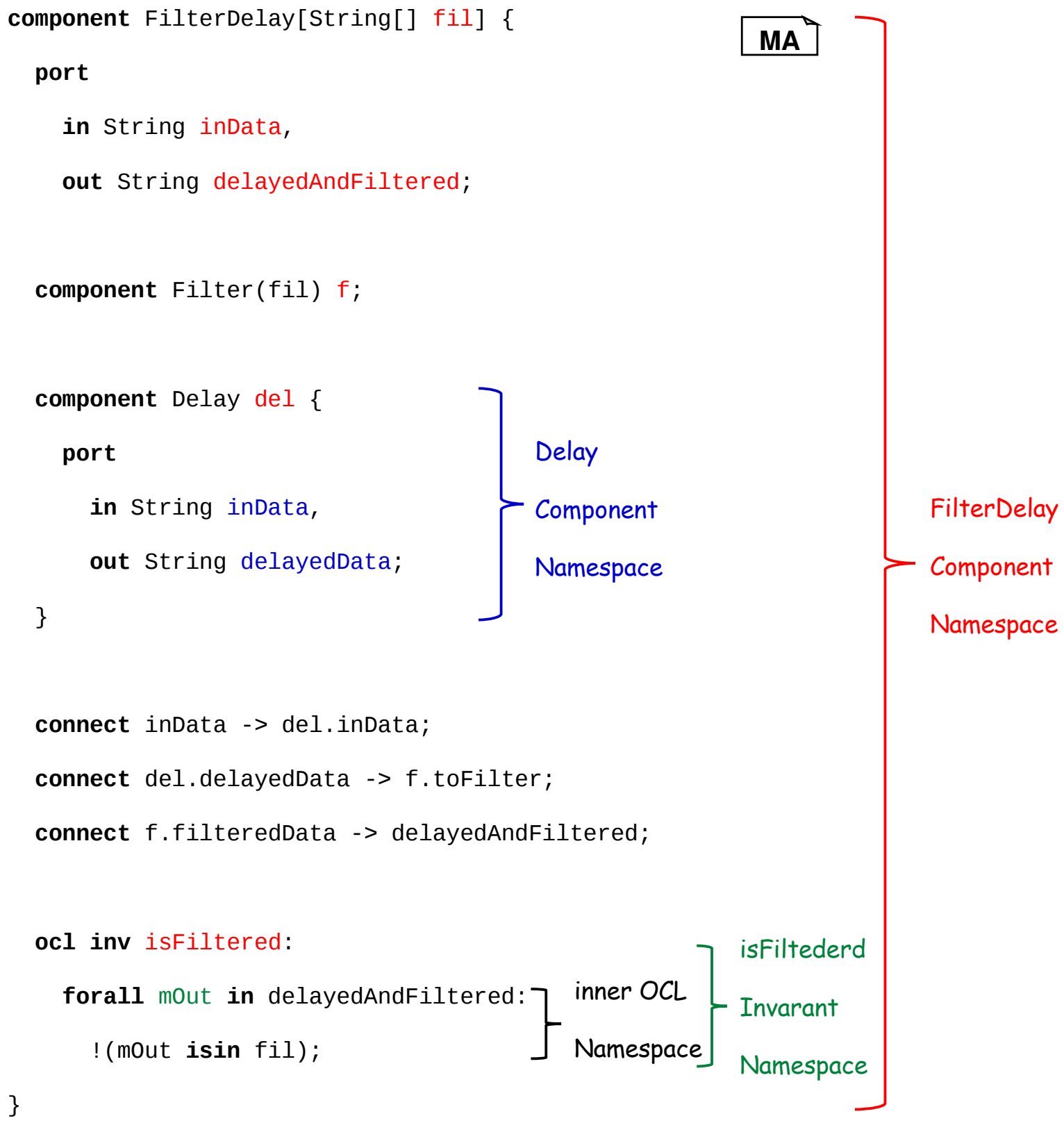

Figure 3.1: Namespaces and identifier declarations in MontiArc.

Please note that using the optional instance name while defining an inner component type will automatically declare a subcomponent with the used instance name (c.f. Sect. 2.3). As this subcomponent is declared in the parent component of the inner component definition, its identifier also belongs to the parents namespace.

The parent namespace `FilterDelay` contains another component namespace that belongs to the inner component type definition `Delay`. All identifiers within this namespace are colored blue. The port name `inData` is still unique, as identifiers of the parent namespace, that also contains this name, are not imported.

Namespaces of invariants import identifiers of their parent namespace, thats why the port name `delayedAndFiltered` as well as the parameter name `fil` may be used inside the namespace of invariant `isFiltered`. It also has a hierarchical structure denoted by the `inner OCL` namespace, as a `forall` construct opens a new namespace in the OCL language.

### B1: All names of model elements within a component namespace have to be unique.

To clearly identify each model element, all names within a component namespace have to be unique. This holds for port names, subcomponent declaration names, generic type parameter names, configuration parameter names, and names of invariants. Listing 3.2 contains several violations of this condition. First, configuration parameter `fil` has the same name as one incoming port (see ll. 1, 4). Second, the subcomponent declaration `del` and an invariant have the same name (c.f. ll. 10, 16).

MontiArc

```
1  component FilterDelay[String[] fil] {
2
3    port
4      in char[][] fil, // 'fil' already declared in l. 1
5      in String inData,
6      out String delayedAndFiltered;
7
8    component Filter(fil) f;
9
10   component Delay del {
11     port
12       in String inData,
13       out String delayedData;
14   }
15
16   ocl inv del: // 'del' already declared in l. 10
17     forall mOut in delayedAndFiltered:
18       !(mOut isin fil);
19 }
```

Listing 3.2: B1: Violation of contxt condition U by using names more then once in a namespace.

### B2: Root component type definitions do not have instance names.

The optional instance name of component type definitions (c.f. Sect. 2.3, page 13) is used to create a subcomponent declaration along with the definition of an inner component type. The created subcomponent then belongs to the parent component type. Root component types do not have a parent, and therefore using an instance name for a root component type definition will result in a not assignable subcomponent. Hence, the usage of instance names for root component definitions is forbidden.

MontiArc

```
component ABPSenderComponent mySenderComp { // instance
                                            // name for
  component Sender innerSender {            // root def.
    // ...                                  // forbidden
  }
  // ...
}
```

Listing 3.3: B2: Instance names of component definitions.

In Lst. 3.3, the component definition `ABPSenderComponent<T>` has an instance name `mySenderComp` which is not allowed. For the inner component definition `Sender` this concept is used to create a subcomponent declaration named `innerSender` along with the definition of the inner component type.

## 3.2 Connections

### CO1: Connectors may not pierce through component interfaces.

MontiArc

```
component A_B_Filter {
  port
    in String msgIn,
    out String msgOut;

  component Filter('a') af;
  component Filter('b') bf;

  connect msgIn -> af.msgs;
  connect bf.filteredMsgs -> msgOut;
  connect af.filteredMsgs -> bf.msgs.d; // d not visible
}

component Filter[char f] {
  port
    in String msgs,
    out String filteredMsgs;
  component Delay(1) d;
  // ...
}
```

Listing 3.4: CO1: Qualified sources and targets of connectors.

Qualified sources or targets of a connector consist of two parts. The first part is a name of a subcomponent, the second part is a port name. Lst. 3.4 contains the definition of component types `A_B_Filter` and `Filter`. The former contains two subcomponent declarations of the latter. The connector show in l. 9 connects port `msgIn` of `A_B_Filter` with port `msgs` of subcomponent `af`. As this port is part of the interface of `af`'s type, this connector is valid. The same holds for the second connector in l. 10 that connects the output of `bf` to the outgoing port `msgOut`. However, the target of the third connector shown in l. 11 is subcomponent `d` that belongs to component type `Filter` that is declared as subcomponent `bf`. As subcomponent declarations are encapsulated and may only be accessed indirectly via their connected ports, `d` is not visible in the scope of `A_B_Filter` and must not be used as a target of a connector.

**CO2: A simple connector's source is an outgoing port of the referenced component type and is therefore not qualified.**

MontiArc

```
1  component A_B_Filter {
2    port
3      in String msgIn,
4      out String msgOut;
5
6    component Filter('a') af
7      [filteredMsgs -> bf.msgs];
8    component Filter('b') bf
9      [bf.filteredMsgs -> bf.msgs]; // source is qualified
10   // ...
11 }
```

Listing 3.5: CO2: Correct and invalid sources of simple connectors.

A source of a simple connector always has to be an outgoing port of the subcomponent's component type. A qualification is therefore not needed as the port is implicitly qualified using the bounded subcomponents name. The first simple connector in line 7 of Lst. 3.5 connects outgoing port `filteredMsgs` of subcomponent `af` with the incoming port `msgs` of subcomponent `bf` and is valid. The source of the second connector in l. 9 contains the subcomponent's name `bf` as an additional qualifier and is therefore invalid.

```
MontiArc
1  component A_B_Filter {
2    port
3      in String msgIn,
4      out String msgOut;
5
6    component Filter('a') af;
7    component Filter('b') bf;
8
9    connect msgIn -> af;
10   connect af -> bf;
11   connect bf -> msgOut;
12 }
```

Listing 3.6: CO3: Using unqualified sources and targets in connectors.

**CO3: Unqualified sources or targets in connectors either refer to a port or a subcomponent declaration.**

If sources or targets of a connector are unqualified, then they must refer to a port or a subcomponent name declared in the scope of the current component type definition. If a name of a subcomponent is used, all yet unconnected ports are connected that have a valid type. For example the first connector given in Lst. 3.6 in l. 9 automatically resolves incoming port `msgs` of subcomponent `af` as the target of the connector, as its type fits to the type of the connector's source. The second connector given in l. 10 connects all compatible outgoing ports of subcomponent `af` with all compatible incoming ports of subcomponent `bf`. Finally, the third connector in l. 11 connects one compatible outgoing port of subcomponent `bf` with the outgoing port `msgOut`. This, however, is only possible if a unique compatible port can be resolved. If more then one compatible port is found, no connections are created and a warning is emitted.

## 3.3 Referential Integrity

**R1: Each outgoing port of a component type definition is used at most once as target of a connector.**

In MontiArc the sender of a message or signal is always unique for the receiver. Hence every receiving port only receives signals from a unique sender, while a sender may transmit its data to more then one receiver. Therefore outgoing ports of a component type definition are used at most once as a target of a connector. In Lst. 3.7 the component type definition `A_B_Filter` violates this condition. The outgoing port `msgOut` is used

MontiArc

```
1 component A_B_Filter {
2   port
3     in String msgIn,
4     out String msgOut;
5
6   component Filter('a') af
7     [filteredMsgs -> msgOut]; // ambiguous sender
8   component Filter('b') bf;
9
10  connect msgIn -> af.msgs, bf.msgs;
11  connect bf.filteredMsgs -> msgOut; // ambiguous sender
12 }
```

Listing 3.7: R1: Unique receivers of connectors.

as a target of the simple connector given in l. 7 and also as a target of the connector given in l. 11. A unique sender may not be identified, as it may be the outgoing port of subcomponent `af` or the outgoing port of subcomponent `bf`. In contrast, a sender of a connector may transmit its messages to more than one receivers. Hence the connector given in l. 10 is valid.

### R2: Each incoming port of a subcomponent is used at most once as target of a connector.

MontiArc

```
1 component A_B_Filter {
2   port
3     in String msgIn,
4     out String msgOut;
5
6   component Filter('a') af;
7   component Filter('b') bf;
8
9   connect msgIn -> bf.msgs,  af.msgs; // ambiguous sender
10  connect bf.filteredMsgs -> af.msgs; // ambiguous sender
11  connect af.filteredMsgs -> msgOut;
12 }
```

Listing 3.8: R2: Unique receivers of connectors.

As already discussed in the previous context condition, the sender of a message is always unique for a receiver. Incoming ports of subcomponents may be used as receivers in a connector and must therefore be used at most once as a receiver in the context of a component type definition. In Lst. 3.8 this context condition is injured by the connectors given in ll. 9–10. The incoming port `msgs` of subcomponent `af` is used twice as a target.

### R3: Full qualified subcomponent types exist in the named package.

MontiArc

```
1 component A_B_Filter {
2   // ...
3   component ma.msg.Filter('a') af;
4   // ...
5 }
```

Listing 3.9: R3: Qualified subcomponent types.

If a qualified component type is used for a subcomponent, a component type definition has to exist in the denoted package. For example the subcomponent declaration shown in Lst. 3.9 uses the qualified type `ma.msg.Filter` (c.f. l. 3). Hence a component definition `Filter` has to exist in package `ma.msg`.

### R4: Unqualified subcomponent types either exist in the current package or are imported using an import statement.

MontiArc

```
1 package ma;
2 import ma.msg.Filter;
3 component A_B_Filter {
4   // ...
5   component Filter('a') af;
6   component C_D_Filter cdf;
7   // ...
8 }
```

Listing 3.10: R4: Unqualified subcomponent types.

If an unqualified component type is used for a subcomponent, it must either exist in the current package or it must be imported using an import statement. Subcomponent `af` given in Lst. 3.10 uses the unqualified type `Filter` that is imported in l. 2. The type of subcomponent `cdf` (l. 6) is unqualified and not imported. Therefore a component type definition `C_D_Filter` has to exist in the current package `ma` given in l. 1.

### R5: The first part of a qualified connector's source or target must correspond to a subcomponent declared in the current component definition.

```
MontiArc
1  component A_B_Filter {
2    port
3      in String msgIn,
4      out String msgOut;
5
6    component Filter('a') af;
7
8    connect msgIn -> af.msgs;
9    connect bf.filteredMsgs -> msgOut;  // subcomponent bf
10 }                                     // does not exist
```

Listing 3.11: R5: Subcomponents in qualified connector parts.

If a source or target of a connector is qualified, the qualifier must be the name of a subcomponent that is declared in the namespace of the current component definition. In Lst. 3.11 the target of the first connector (l. 8) is qualified with `af`. As a subcomponent `af` is declared in l. 6 the qualifier is valid. In contrast the source of the second connector (l. 9) is qualified with `bf`, but a subcomponent with that name is not declared. Hence, this connector is invalid.

### R6: The second part of a qualified connector's source or target must correspond to a port name of the referenced subcomponent determined by the first part.

The second part of a qualified source or target of a connector is a port name. A port with that name must exist in the component type of the subcomponent that is given by the qualifier. In Lst. 3.12 the target of the first connector given in l. 12 is port `toDelay` of subcomponent `del`. As shown in l. 8 the component type of this subcomponent contains this port. Hence, the first connector is valid. The source of the second connector (l. 13) is port `delayed` of subcomponent `del`. As this port does not exist in component type `Delay` (c.f. ll. 6–10), this connector is invalid.

### R7: The source port of a simple connector must exist in the subcomponents type.

In simple connectors, the source directly references an outgoing port in the type of the subcomponent to which the simple connector belongs to. This

MontiArc

```
component FilterDelay {
  port
    in String inputData,
    out String delayed;

  component Delay del {
    port
      in String toDelay,
      out String delayedData;
  }

  connect inputData -> del.toDelay;
  connect del.delayed -> delayed;  // port delayed does
}                                  // not exist
```

Listing 3.12: R6: Ports in qualified connector parts.

MontiArc

```
1  component FilterDelay {
2    port
3      out String delayed1,
4      out String delayed2;
5
6    component Delay {
7      port
8        in String toDelay,
9        out String delayedData;
10   }
11
12   component Delay
13     d1 [delayedData -> delayed1],
14     d2 [delayed -> delayed2];  // port delayed does
15 }                              // not exist
```

Listing 3.13: R7: Sources of simple connectors.

port has to exist. In Lst. 3.13 the source of the first simple connector in l. 13 exists and the connector is therefore valid. As the component type `Delay` does not have an outgoing port `delayed` (c.f. ll. 6–10), the second simple connector given in l. 14 is invalid.

### R8: The types of two connected ports have to be compatible, i.e., the target port has the same type or is a supertype of the source port type.

To assure type correct communication, source and target ports of connectors have to be compatible. A receiver may be connected to a sender, if both

MontiArc

```
1  component MyComp {
2    port
3      in Integer myInt,
4      out Object myObj;
5
6    component Buffer<Integer> bInt;
7    component Buffer<Object> bObj;
8    component Buffer<String> bStr;
9
10   connect myInt -> bInt.input;
11   connect bInt.buffered -> bObj.input;
12   connect bObj.buffered -> bStr.input; // incompatible
13   connect bStr.buffered -> myObj;       // types Object,
14 }                                       // String
15 component Buffer<T> {
16   port
17     in T input,
18     out T buffered;
19 }
```

Listing 3.14: R8: Type compatible connectors.

have the same type or the receiver type is a supertype of the source type. Lst. 3.14 contains some examples for connectors with different source and target types. The first connector in l. 10 ist obviously valid, as source and target type are the same. The second connector in l. 11 connects a source port with type `Integer` and a target port with type `Object`. As `Object` is a supertype of `Integer`, this connection is valid. The third connector (l. 12) connects `Object` with `String`. Because `String` is a subtype of `Object` and not a supertype, it is invalid. The fourth connector in l. 13 is valid again, as `Object` is a supertype of `String`.

### R9: If a generic component type is instantiated as a subcomponent, all generic parameters have to be assigned.

A generic component is a component that defines generic type parameters in its head (see Sect. 2.1, page 12). If such a component type is used as a subcomponent type, a data type has to be assigned to each of these generic type parameters. Lst. 3.15 contains the definition of the generic component type `Buffer` in ll. 1–5 that has two generic type parameters `K` and `V`. In the component type definition `MyComp` in ll. 6–11 two subcomponents are declared that have the aforementioned type. The first subcomponent declaration (l. 8.) assigns a data type to each type parameter and is valid. The incoming port `input` of `b1` has now the type `Integer`, the outgoing port has the type `String`. The second subcomponent declaration `b2` in l. 9

MontiArc

```
component Buffer<K, V> {
  port
    in K input,
    out V buffered;
}
component MyComp {
  // ...
  component Buffer<Integer, String> b1;
  component Buffer<Integer> b2;  // type parameter V
  // ...                           // not assigned
}
```

Listing 3.15: R9: Using generic component types as subcomponent types.

only assigns one type parameter. As the `Buffer` component type claims two generic type parameters, the subcomponent declaration is invalid.

### R10: If a configurable component is instantiated as a subcomponent, all configuration parameters have to be assigned.

MontiArc

```
1  component LossyDelay<T>[int delay, int lossrate] {
2    port
3      in T msgIn,
4      out T delayed;
5  }
6  component MyComp {
7    // ...
8    component LossyDelay<String>(1, 5) ld1;
9    component LossyDelay<String>(1) ld2;  // missing
10   // ...                                   // parameter
11 }                                          // lossrate
```

Listing 3.16: R10: Using configurable component types as subcomponent types.

A configurable component defines configuration parameters in its head (see Sect. 2.1, page 12). If such a component type is used as a subcomponent type, a value has to be assigned to each configuration parameter. In Lst. 3.16 the configurable component type `LossyDelay` defined in ll. 1–5 is used as type of subcomponent `ld1` in l. 8. In the subcomponent declaration a value is assigned to both configuration parameters. Therefore the subcomponent declaration is valid. The second subcomponent declaration in l. 9 only assigns one value, as two values are expected, the declaration is invalid.

### R11: Inheritance cycles for components are forbidden.

```
MontiArc
1 component ABPReceiver<T>
2     extends CommonReceiver<T> { // inheritance cycle
3   // ...
4 }
5
6 component CommonReceiver<T>
7     extends ABPReceiver<T> { // inheritance cycle
8   // ...
9 }
```

Listing 3.17: R11: An inheritance cycle in MontiArc.

Lst. 3.17 shows an example for an inheritance cycle. The component type `ABPReceiver` extends the `CommonReceiver` component type (l. 2) which is a subtype of the `ABPReceiver` component (l. 7). Such a system cannot be instantiated, therefor inheritance cycles are forbidden.

### R12: An inner component type definition does not extend the component type in which it is defined.

```
MontiArc
component Outer {

  component Inner extends Outer { // structural
    // ...                          // inheritance cycle
  }
  // ...
}
```

Listing 3.18: R12: Structural extension cycle.

A structural extension cycle is given, if an inner component type definition extends the component type of its surrounding parent component. As the inner component will import itself in a structural extension cycle, it may not be instantiated using our mechanism. Therefore it is forbidden for inner component type definitions to extend its parent component. This context condition is violated in Lst. 3.18 where the inner component type `Inner` extends its parent component type `Outer`.

```
MontiArc
1 component A {
2   // ...
3   component B myB; // reference cycle
4 }
5 component B {
6   // ...
7   component A myA; // reference cycle
8 }
```

Listing 3.19: R13: Structural extension cycle.

### R13: Subcomponent reference cycles are forbidden.

A reference cycle is given, if two component types declare each other as subcomponets. As instantiation of such a system will result in an endless instantiation process, these cycles are forbidden. An example for a reference cycle is shown in Lst. 3.19. Component type `A` contains a subcomponent declaration of type `B` (c.f. l. 3). The component type `B` contains itself a subcomponent of type `A` (c.f. l. 7). If component type `A` is instantiated, an instance of component type `B` is created that will itself create another instance of `A` and so forth.

## 3.4 Conventions

### CV1: Instance names should start with a lower-case letter.

```
MontiArc
component Inverter<T> [Number delta] {
  port
    in Integer input,
    out Integer inverted;

  component Filter(delta) myFilter;

  java inv isInverted: {
    //...
  }
}
```

Listing 3.20: CV1 and CV2: Naming Conventions of MontiArc

Names in the scope of component definitions should start with a lower case letter. This context condition affects names of ports, subcomponent declarations, configuration parameters, and invariants. Therefor all names contained in the component definition depicted in Lst. 3.20 obey this rule. Violating this context condition will result in a warning.

### CV2: Types should start with an upper-case letter.

Component types and generic type parameters should start with an upper case letter. Hence the component name `Inverter` as well as the used generic type parameter `T` are well formed. Violating this context condition will result in a warning.

### CV3: Duplicated imports should be avoided.

Defining identical imports more than once will result in a warning.

### CV4: Unused direct imports should be avoided.

The definition of imports which are not used in the model violates this convention and results in a warning.

### CV5: In decomposed components all ports should be used in at least one connector.

MontiArc

```
1  component A_Filter {
2    port
3      in String msgIn,
4      in String foo,  // unused port
5      out String msgOut;
6
7    component Filter('a') af;
8
9    connect msgIn -> af.msgs;
10   connect af.filteredMsgs -> msgOut;
11 }
```

Listing 3.21: CV5: Using all ports.

If incoming or outgoing ports of a decomposed component type are not used in at least one connector, a warning is produced to inform the modeler that parts of the components interface are unconnected. In Lst. 3.21 the ports `msgIn` and `msgOut` are both used and connected to subcomponents (c.f. ll. 9-10). In contrast port `foo` is not connected and a warning is produced (l. 4).

**CV6: All ports of subcomponents should be used in at least one connector.**

MontiArc

```
1  component A_Filter {
2    port
3      in String msgIn,
4      out String msgOut;
5
6    component Filter('a') af;
7    component Filter('b') bf;  // unconnected ports msgs,
8                               // filteredMsgs
9    connect msgIn -> af.msgs;
10   connect af.filteredMsgs -> msgOut;
11 }
```

Listing 3.22: CV6: Using all ports of subcomponents.

If ports of subcomponents are unconnected, this may result in an unexpected behavior. Hence, the modeler is informed with a warning, if subcomponents in a decomposed component type definition have unconnected ports. All ports of subcomponent `af` in Lst. 3.22 are connected in the connectors given in ll. 9-10. But no ports of subcomponent `bf` are connected, therefore a warning is created.

# Chapter 4

# Semantics

A complete modeling language definition consists not only of the syntax of the language but also of the language's semantics in (terms of meaning [HR04]). We have given some examples of models of the MontiArc language in Chapters 1-3 and intuitively explained the semantics of the models presented. When working with modeling languages, building models, and developing tools the models' semantics requires a more formal definition than a verbal and often vague description as given above. A formal and sound background is a prerequisite for building integrated model-based development frameworks [BS01, BR07].

In general the semantics of a language $L$ in terms of its meaning expressed in a semantic domain $S$ can be defined using a semantic mapping function $m : L \to \mathcal{P}(S)$ as illustrated in Fig. 4.1 (see [HR04]). The idea behind this denotational semantics is simply to *explain* every model of $L$ in terms of the semantic domain.

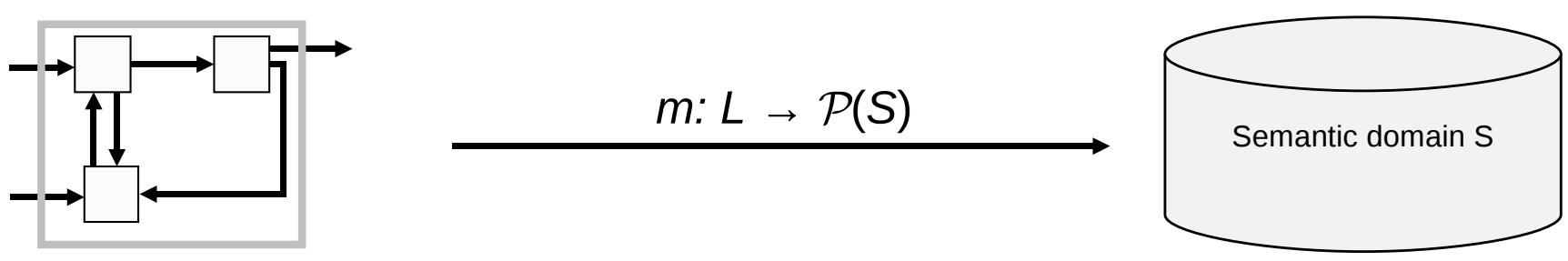

Figure 4.1: Semantic mapping of a modeling language to its semantic domain

The semantic domain $S$ is a formal model with well defined and well understood elements, which is able to capture the meaning of elements of systems modeled using a modeling language. In the general case, where models can be under-specified or do not have an exact one-to-one representation in the semantic domain, one model (as an element of the language $L$) can be mapped to a set of elements of the semantic domain that capture its meaning – including alternatives due to the underspecification. If the domain of stream processing functions $SPF$ (see [BDD$^+$93, Rum96, RR11]) was chosen to represent MontiArc components, the semantics of the Mon-tiArc component `LossyChannel` from Fig. 4.2 could be the set of all stream processing functions with the (simplified) signature $f : \mathbb{Z}^\omega \to \mathbb{Z}^\omega$ since nothing is known about the behavior of the `LossyChannel` component from looking at the MontiArc model itself. A combination of structure and behavior in the semantics domain allows for example the refinement and evolution of architectures [PR97, PR99].

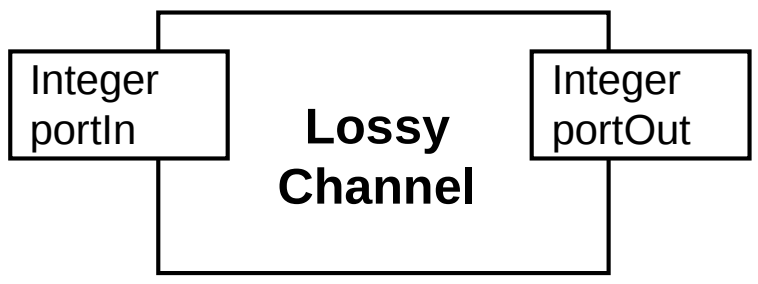

Figure 4.2: The component `LossyChannel` (from the example in Sect. 1.1)

In the following semantics definition we focus on the structure of systems modeled using the ADL MontiArc. It is however no problem to extend the semantic domain presented in Sect. 4.1 to a semantic domain that supports also behavior as demonstrated in [BCR06, BCR07a, BCR07b, Grö10] for object oriented systems. An extension of the semantic domain and mapping function with behavior is straight forward as shown and discussed in [BDD$^+$93, BS01, BR07]. MontiArc can be seen as one of the modular system views discussed in [BR07] describing interfaces, hierarchical decomposition, and composition of components.

The semantic domain we present in the next section makes the semantic mapping of MontiArc models a function $m : L \to S$, that maps every well formed model to *exactly one element* of the semantic domain.

## 4.1 A Semantic Domain for MontiArc

We give a semantic domain that captures the structure of systems described using the ADL MontiArc. It consists of components with ports, subcomponents, and connectors as shown in the following equations and is derived from the grammars of the MontiArc language family:

$$\begin{aligned} \text{Component} &= \text{CType} \times \mathcal{P}(\text{Port}) \times \mathcal{P}(\text{SubComponent}) \times \\ &\quad\ \mathcal{P}(\text{Connector}) \qquad (4.1) \\ \text{Port} &= \{\text{IN}, \text{OUT}\} \times \text{PType} \times \text{PName} \qquad (4.2) \\ \text{SubComponent} &= \text{CName} \times \text{Component} \qquad (4.3) \\ \text{Connector} &= \text{CName} \times \text{PName} \times \text{CName} \times \text{PName} \qquad (4.4) \end{aligned}$$

The notation $\mathcal{P}(\text{Port})$ refers to the power set of the set Port. The sets CType of component type names, CName of subcomponent names, PName of port names, and PType of port data types are not further specified and could, e.g., consist of simple strings as well as possibly more complex elements defining their own hierarchy like the Java type system.

The construct Component of equation 4.1 represents components with their component type, ports, subcomponents, and connectors. Please not that this semantic domain does not necessarily require component types since these are reflected in the set of ports, subcomponents and connectors. We chose to add component types to the semantic domain as an extension to allow distinguish for example a `Filter` from a `Delay` component which might have identical interfaces and both no further subcomponents. Also the set CType might well support generic component types, e.g., by a string representation of the complete parametrized type name.

Connectors (equation 4.4) are part of their owning component and reference the component's subcomponents and ports by their names. For convenience we require that the set CName contains an element † that is similar to the keyword `this` used in some object oriented languages. A connection in the semantic domain from a port of the parent component to a port of one of its children would use the †-symbol to refer to the parent component. For example the MontiArc connector `connect messageIn -> sender.messageIn` maps to the tuple $(\dagger, \text{messageIn}, \text{sender}, \text{messageIn})$ in the semantic domain.

In the following we refer to elements of the structures Component, Port, SubComponent, and Connector by using the abbreviation .cName to refer to a field of type CName of a tuple (if the element position in the tuple is unique). We give some well-formedness rules of the semantic domain, similar to MontiArc context conditions discussed in Chapter 3:
Component types determine the structure of a component:

$$\begin{aligned} \forall c_1, c_2 \in \ & \text{Component} : \\ & c_1.\text{cType} = c_2.\text{cType} \Rightarrow c_1 = c_2 \end{aligned} \tag{4.5}$$

Please note that on the other hand an identical structure does not imply the same component type.
The names of all ports of a component are unique:

$$\begin{aligned} \forall c \in \ & \text{Component}, p_1, p_2 \in c.\text{ports} : \\ & p_1.\text{pName} = p_2.\text{pName} \Rightarrow p_1 = p_2 \end{aligned} \tag{4.6}$$

The names of all subcomponents of a component are unique:

$$\begin{aligned} \forall c \in \ & \text{Component}, sc_1, sc_2 \in c.\text{subComponents} : \\ & sc_1.\text{name} = sc_2.\text{name} \Rightarrow sc_1 = sc_2 \end{aligned} \tag{4.7}$$

Connectors can only connect ports of the component they belong to or its direct subcomponents where the ports have the correct direction:

$$\forall c \in \text{Component}, \forall (s, pn_s, r, pn_r) \in c.\text{connectors} :$$

$$
\begin{aligned}
&(s = \dagger \Rightarrow \exists p \in c.\text{ports} : p.\text{pName} = pn_s \wedge p.\text{direction} = IN) \wedge \\
&(s \neq \dagger \Rightarrow \exists (s, c') \in c.\text{subComponents}, p \in c'.\text{ports} : \\
&\quad p.\text{pName} = pn_s \wedge p.\text{direction} = OUT) \wedge \\
&(r = \dagger \Rightarrow \exists p \in c.\text{ports} : p.\text{pName} = pn_r \wedge p.\text{direction} = OUT) \wedge \\
&(r \neq \dagger \Rightarrow \exists (r, c') \in c.\text{subComponents}, p \in c'.\text{ports} : \\
&\quad p.\text{pName} = pn_r \wedge p.\text{direction} = IN) \qquad (4.8)
\end{aligned}
$$

Every port reads at most from one port connected by a unique connector:

$$
\begin{aligned}
\forall c \in\ & \text{Component}, \forall (s1, pn1_s, r1, pn1_r), (s2, pn2_s, r2, pn2_r) \in c.\text{connectors} : \\
& r1 = r2 \wedge pn1_r = pn2_r \Rightarrow \\
& \quad (s1, pn1_s, r1, pn1_r) = (s2, pn2_s, r2, pn2_r) \qquad (4.9)
\end{aligned}
$$

The domain given in equations (4.1) - (4.4) is a simplified and abstract version of systems modeled using the MontiArc language. The list of rules (4.5) - (4.9) is not complete but demonstrates the most important properties of the semantic domain. This formalization of the essential concepts helps when reasoning about the meaning of different models.

## 4.2 A Semantic Mapping for MontiArc

We sketch a semantic mapping of MontiArc to the semantic domain introduced in the previous section to highlight some important decisions in the design of MontiArc's semantics.

A complete and formal definition of the semantic mapping function $m : L \rightarrow S$ can be done compositionally as, for example, shown in [CGR08] for class diagrams and sketched in [GRR09] for general modeling languages: When defining the semantic mapping of a model, children elements of nodes in the abstract syntax tree can be mapped by independent or adequately parametrized mapping functions thus simplifying the definition of the mapping by decomposition.

Here we only show some examples to illustrate the semantic mapping. The result of mapping the `LossyChannel` component from Fig. 4.2 to its semantics is as expected:

$$
\begin{aligned}
\text{Component} &= \{(\text{LossyChannel}, \text{ports}, \emptyset, \emptyset)\} \\
\text{ports} &= \{(\text{IN}, \text{Integer}, \text{portIn}), \\
&\quad\ (\text{OUT}, \text{Integer}, \text{portOut})\} \\
\text{Port} &= \text{ports} \\
\text{SubComponent} &= \emptyset \\
\text{Connector} &= \emptyset
\end{aligned}
$$

$$
\begin{aligned}
\text{CType} &= \{\text{LossyChannel}\}\\
\text{PType} &= \{\text{Integer}\}\\
\text{PName} &= \{\text{portIn}, \text{portOut}\}\\
\text{CName} &= \emptyset
\end{aligned}
$$

A more interesting example is the semantics of the structured component `BoardLightsControl` given in Lst. 4.3. In this example the component definitions `TurnSignalController` and `HeadLightsController` are referenced as subcomponents.

```
 1 package automotive.ecu;
 2
 3 import automotive.ecu.controller.TurnSignalController;
 4 import automotive.ecu.controller.HeadLightsController;
 5
 6 component BoardLightsControl {
 7
 8     autoconnect port;
 9
10     port
11         /* ... */
12
13     component TurnSignalController frontSignalController;
14
15     component TurnSignalController rearSignalController;
16
17     component HeadLightsController;
18 }
```

Listing 4.3: The component `BoardLightsControl` reusing the component `TurnSignalController` twice as `frontSignalController` and `rearSignalController`

The corresponding part of the semantics of component `BoardLightsControl` is given in the following equations:

$$
\begin{aligned}
\text{Component} &= \{(\text{BoardLightsControl}, \{\ldots\}, \text{subComps}_{BLC}, \{\ldots\}),\\
&\quad (\text{TurnSignalController}, \{\ldots\}, \{\ldots\}, \{\ldots\}) = TSC,\\
&\quad (\text{HeadLightsController}, \{\ldots\}, \{\ldots\}, \{\ldots\}) = HLC,\\
&\quad \ldots\}\\
\text{subComps}_{BLC} &= \{(\text{frontSignalController}, TSC),\\
&\quad (\text{rearSignalController}, TSC),\\
&\quad (\text{headLightsController}, HLC)\}
\end{aligned}
$$

$$
\begin{aligned}
\text{SubComponent} &= \text{subComps}_{BLC} \cup \{\ldots\} \\
\text{Port} &= \{\ldots\} \\
\text{Connector} &= \{\ldots\} \\
\text{CType} &= \{\text{BoardLightsControl}, \text{TurnSignalController}, \\
&\quad\ \text{HeadLightsController}, \ldots\} \\
\text{PType} &= \{\ldots\} \\
\text{CName} &= \{\text{frontSignalController}, \text{rearSignalController}, \\
&\quad\ \text{headLightsController}, \ldots\} \\
\text{PName} &= \{\ldots\}
\end{aligned}
$$

The set Component consists of all components in the semantics of the model. Its first element represents the component type `BoardLights-Control`. By containing set $\text{subComps}_{BLC}$ it includes besides the subcomponent headLightsController (with structure $HLC$) the subcomponents frontSignalController and rearSignalController both of type `TurnSignal-Controller`. These subcomponents have identical structure (abbreviated as $TSC$) in the semantics.

In contrast to the MontiArc syntax level – where these components were just defined once and then referenced – in the semantic domain they are identical copies of the referenced component. This unfolding of the nested references into components makes the structure of an architecture more explicit.

## 4.3 Semantic Mapping Applications

In Chapter 2 we have introduced the language hierarchy of MontiArc starting from the basic ADL ArcD. For this language we have developed a code generator that translates the modeled architectures into code that can be executed within a simulation framework. Furthermore we have developed tool support for context condition checks, a symbol table infrastructure, and an Eclipse editor.

MontiArc extends the language ArcD and reuses most of the developed infrastructure. One specific case of reuse is done by transforming models of the syntactically richer language MontiArc to equivalent ArcD models. This is done by a set of transformations on the models' abstract syntax. One of these transformations is, e.g., replacing the `autoconnect` statement by a set of connectors added to the abstract syntax. These transformations are implemented in Java and assumed to be semantics preserving. The MontiArc keyword `autoconnect` is removed and the connectors added to the abstract syntax are concepts inherited from ArcD and their code generation is already implemented in the ArcD code generator.

To prove the correctness of these transformations one could define a semantic mapping $m$ from the language MontiArc containing the additional language features to the semantic domain presented above. Applying the semantic mapping $m$ to a MontiArc model should yield the same result as first applying the AST transformations and then the mapping $m_{simp}$ from the simpler language (see Fig. 4.4).

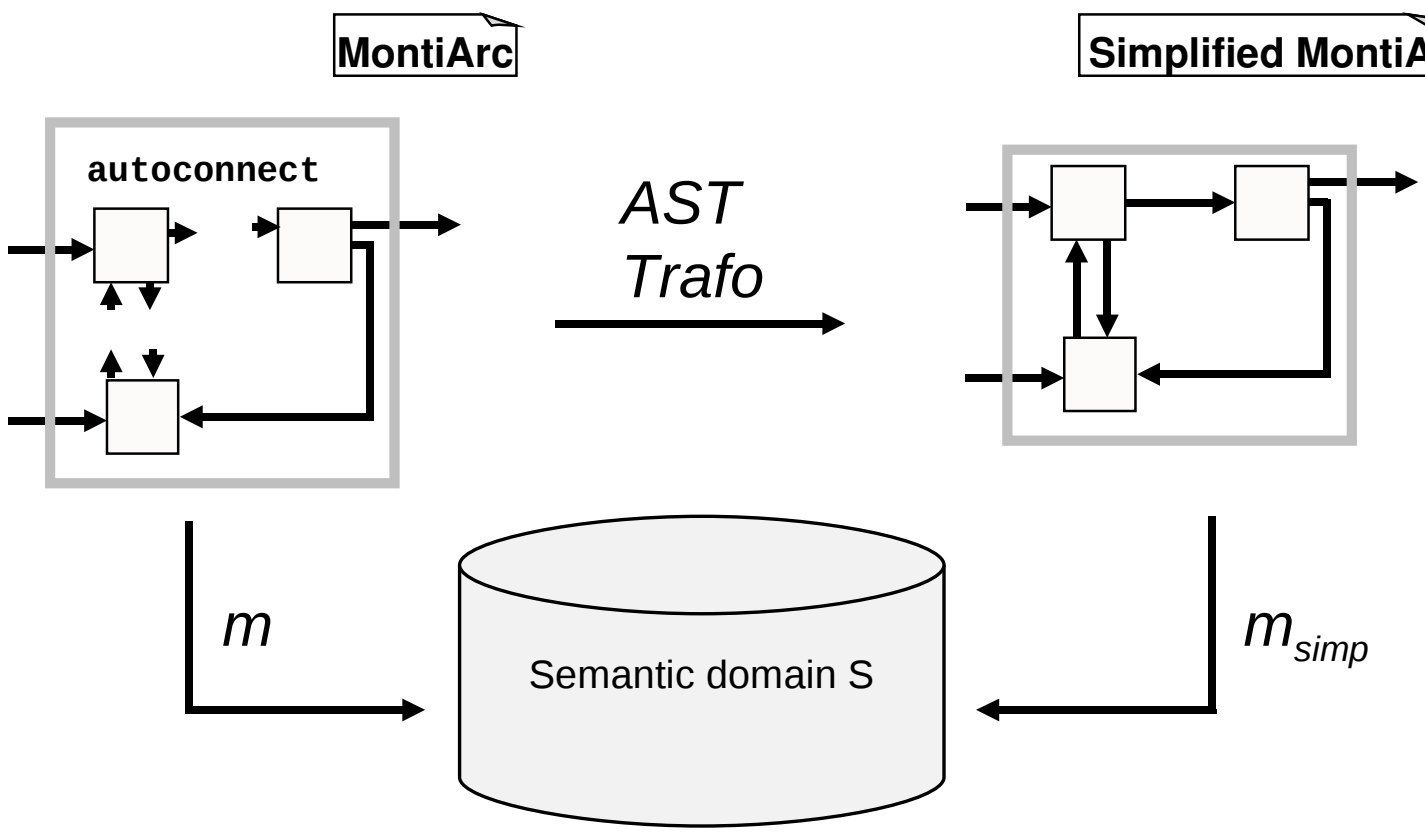

Figure 4.4: Semantic mapping of MontiArc and simplified MontiArc to show correctness of AST transformations

With a complete definition of the abstract syntax, the semantic domain, and the semantic mappings $m$ and $m_{simp}$ one could apply the semantic mappings to the MontiArc model and the transformed ArcD model to show that the AST transformation is semantics preserving. With an additional formalization of the AST transformation one could even show the correctness of the transformation for all MontiArc models:

$$\forall l \in \text{MontiArc} : m_{simp}(\text{astTrafo}(l)) = m(l).$$

# Appendix A

# Simplified Grammars for Humand Reading

These grammars are meant for human reading. They are MontiArc grammars, where all parsing pragmas have been striped of. They still describe the full context free language.

## A.1 Readable Architectural Diagrams Grammar

Readable Grammar for ArcD

```
package mc.umlp.arcd;

/**
 * Simplified ArcD grammar.
 *
 * @author Arne Haber
 */
grammar SimpleArchitectureDiagram {

/* ====================================================*/
/* ====================== OPTIONS =====================*/
/* ====================================================*/
  options {
    compilationunit ArcComponent
  }

/* ====================================================*/
/* ==================== PRODUCTIONS ===================*/
/* ====================================================*/
  /**
   * A component may contain arbitrary many ArcElements.
   * This interface may be used as an extension point to
```

```
23  * enrich components with further elements.
24  */
25  interface ArcElement;
26
27  /**
28  * A component is a unit of computation or a data store.
29  * The size of a component may scale from a single
30  * procedure to a whole application. A component may be
31  * either decomposed to subcomponents or is atomic.
32  */
33  ArcComponent implements ArcElement =
34    Stereotype?
35    "component" Name (instanceName:Name)?
36    ArcComponentHead ArcComponentBody;
37
38  /**
39  * A components head is used to define generic type
40  * parameters that may be used as port types in the
41  * component, to define configuration parameters that may
42  * be used to configure the component, and to set the
43  * parent component of this component.
44  */
45  ArcComponentHead =
46    TypeParameters?
47    ("[" ArcParameter* "]")?
48    ("extends" ReferenceType)?;
49
50  /**
51  * The body contains architectural elements of
52  * this component.
53  */
54  ArcComponentBody =
55    "{"
56      ArcElement*
57    "}";
58
59  /**
60  * An ArcInterface defines an interface of a component
61  * containing in- and outgoing ports.
62  */
63  ArcInterface implements ArcElement =
64    Stereotype?
65    "port" (ArcPort)* ";";
66
67  /**
68  * An incoming port is used to receive messages, an
69  * outgoing port is used to send messages of a specific
70  * type.
71  */
```

```
ArcPort =
  Stereotype?
  ("in" | "out")
  Type Name?;

/**
* A subcomponent is used to create one or more instances
* of another component. This way the hierarchical
* structure of a component is defined.
*
*/
ArcSubComponent implements ArcElement =
  Stereotype?
  "component"
  ReferenceType
  ("(" ArcConfigurationParameter* ")" )?
  (ArcSubComponentInstance* )? ";";

/**
* A subcomponent instance binds the name of an instance
* with an optional list of simple connectors used to
* connect this instance with other subcomponents/ports.
*/
ArcSubComponentInstance =
  Name
  ("[" ArcSimpleConnector
  (";" ArcSimpleConnector)* "]")?;

/**
* A connector connects one source port with one or many
* target ports.
*/
ArcConnector implements ArcElement=
  Stereotype?
  "connect" source:QualifiedName "->"
  targets:QualifiedName ("," targets:QualifiedName)* ";";

/**
* A simple way to connect ports.
*/
ArcSimpleConnector =
  source:QualifiedName "->" targets:QualifiedName
                       ("," targets:QualifiedName)*;

/**
* ArcParameters are used in configurable components.
*/
ArcParameter =
  Type Name;
```

```
121
122  /**
123  * ArcConfigurationParameter are used to configure
124  * configurable components. It is either a value or a
125  * variable name.
126  */
127  ArcConfigurationParameter =
128    QualifiedName | Literal;
129 }
```

Listing A.1: Simplified common MontiCore grammar for architectural diagrams

## A.2 Readable MontiArc Grammar

Readable Grammar for MontiArc

```
package mc.umlp.arc;

/**
 * Simplified MontiArc grammar.
 */
grammar SimpleMontiArc extends ArchitectureDiagram {

/* ======================================================*/
/* ====================== OPTIONS =====================*/
/* ======================================================*/
  options {
    compilationunit ArcComponent
  }

/* ======================================================*/
/* =================== PRODUCTIONS ===================*/
/* ======================================================*/

  /**
   * MontiArc components may contain arbitrary many
   * configurations. These configurations have to
   * implement this interface.
   */
  interface MontiArcConfig;

  /**
   * Extends the component body from super grammar with a
   * configuration.
   * @Overwrite ArchitectureDiagram.ArcComponentBody
   */
  ArcComponentBody =
    "{"
      MontiArcConfig*
      ArcElement*
    "}";

  /**
   * An invariant constrains the behavior of a component.
   */
  MontiArcInvariant implements ArcElement =
    "inv" Name ":" InvariantContent ";";

  /**
   * AutoConnect is used to connect ports automatically.
   */
```

```
46   MontiArcAutoConnect implements MontiArcConfig =
47     "autoconnect" Stereotype?
48     ("type" | "port" | "off") ";";
49
50   /**
51   * Autoinstantiate is used to instantiate inner components
52   * without generic parameters or configuration parameters
53   * automatically. If more then one instance of this inner
54   * component is created by using a reference, the
55   * automatically instanciated reference will dissapear.
56   */
57   MontiArcAutoInstantiate implements MontiArcConfig  =
58     "autoinstantiate" Stereotype?
59     ("on" | "off") ";";
60
61   /**
62   * Sets the time behaviour from the component.
63   */
64   MontiArcTimingParadigm implements MontiArcConfig  =
65     "behavior" Stereotype?
66     ("timed" | "untimed" | "timesynchronous") ";";
67 }
```

Listing A.2: Simplified MontiCore grammar for MontiArc

# Appendix B

# Complete Grammars for Parsing

These grammars are the exact versions used in the MontiCore framework to define tool support.

## B.1 Architectural Diagrams Grammar

Grammar for ArcD

```
package mc.umlp.arcd;

version "1.0";
/**
 * Grammar for common architectural elements. Provides
 * infrastructure for component definitions, component
 * interface definitions, and the hierarchical structure
 * of components.
 *
 * @author Arne Haber
 */
grammar ArchitectureDiagram extends mc.umlp.common.Common {

/* =======================================================*/
/* ====================== OPTIONS =====================*/
/* =======================================================*/
  options {
    compilationunit ArcComponent
    nostring
    parser lookahead=5
    lexer lookahead=7
  }
```

```

/* ==========================================================*/
/* ==================== PRODUCTIONS ====================*/
/* ==========================================================*/
  /**
  * A component may contain arbitrary many ArcElements.
  * This interface may be used as an extension point to
  * enrich components with further elements.
  */
  interface ArcElement;

  /**
  * A component is a unit of computation or a data store.
  * The size of a component may scale from a single
  * procedure to a whole application. A component may be
  * either decomposed to subcomponents or is atomic.
  *
  * @attribute stereotype an optional stereotype
  * @attribute name type name of this component
  * @attribute instanceName if this optional name is given,
  *   it is automatically created a subcomponent that
  *   instantiates this inner component. This is only
  *   allowed for inner component definitions.
  * @attribute head is used to set generic types, a
  *   configuration and a parent component
  * @attribute body contains the architectural elements
  *   inherited by this component
  */
  /ArcComponent implements
      (Stereotype? "component" Name Name? ArcComponentHead
         "{")=> ArcElement =
    Stereotype?
    "component" Name (instanceName:Name)?
    head:ArcComponentHead
    body:ArcComponentBody;

  /**
  * A components head is used to define generic type
  * parameters that may be used as port types in the
  * component, to define configuration parameters that may
  * be used to configure the component, and to set the
  * parent component of this component.
  *
  * @attribute genericTypeParameters a list of type
  *   parameters that may be used as port types in the
  *   component
  * @attribute parameters a list of ArcParameters that
  *   define a configurable component. If a configurable
  *   component is referenced, these parameters have to be
```

```
71  *   set.
72  * @attribute superComponent the type of the super
73  *   component
74  */
75  /ArcComponentHead =
76    (options {greedy=true;}:
77      genericTypeParameters:TypeParameters)?
78    ("[" parameters:ArcParameter
79    ("," parameters:ArcParameter)* "]")?
80    ("extends" superComponent:ReferenceType)?;
81
82  /**
83  * The body contains architectural elements of
84  * this component.
85  *
86  * @attribute arcElement list of architectural elements
87  */
88  /ArcComponentBody =
89    "{"
90      ArcElement*
91    "}";
92
93  /**
94  * An ArcInterface defines an interface of a component
95  * containing in- and outgoing ports.
96  *
97  * @attribute stereotype an optional stereotype
98  * @attribute ports a list of ports that are contained in
99  *   this interface
100 */
101 /ArcInterface implements (Stereotype? "port")=>
       ArcElement =
102   Stereotype?
103   "port" ports:ArcPort ("," ports:ArcPort)* ";";
104
105 /**
106 * An incoming port is used to receive messages, an
107 * outgoing port is used to send messages of a specific
108 * type.
109 *
110 * @attribute stereotype an optional stereotype
111 * @attribute incoming true, if this is an incoming port
112 * @attribute outgoing true, if this is an outgoing port
113 * @attribute type the message type of this port
114 * @attribute name an optional name of this port
115 */
116 /ArcPort =
117   Stereotype?
118   (incoming:["in"] | outgoing:["out"])
```

```
119      Type Name?;
120
121    /**
122    * A subcomponent is used to create one or more instances
123    *  of another component. This way the hierarchical
124    * structure of a component is defined.
125    *
126    * @attribute stereotype an optional stereotype
127    * @attribute type the type of the instantiated component
128    * @attribute arguments list of configuration parameters
129    *   that are to be set, if the instantiated component is
130    *   configurable.
131    * @attribute instances list of instances that should be
132    *   created
133    */
134    /ArcSubComponent implements
135        (Stereotype? "component" ReferenceType
136        ("(" | Name | ";" ))=> ArcElement =
137      Stereotype?
138      "component"
139      type:ReferenceType
140      ("(" arguments:ArcConfigurationParameter
141      ("," arguments:ArcConfigurationParameter)* ")" )?
142      (instances:ArcSubComponentInstance
143      ("," instances:ArcSubComponentInstance)* )? ";";
144
145    /**
146    * A subcomponent instance binds the name of an instance
147    * with an optional list of simple connectors used to
148    * connect this instance with other subcomponents/ports.
149    *
150    * @attribute name the name of this instance
151    * @attribute connectors list of simple connectors
152    */
153    /ArcSubComponentInstance =
154      Name
155      ("[" connectors:ArcSimpleConnector
156      (";" connectors:ArcSimpleConnector)* "]")?;
157
158    /**
159    * A connector connects one source port with one or many
160    * target ports.
161    *
162    * @attribute source source port or component instance
163    *   name
164    * @attribute targets a list of target ports or component
165    *   instance names
166    */
167    /ArcConnector implements
```

```
    (Stereotype? "connect" QualifiedName "->")=>
        ArcElement=
  Stereotype?
  "connect" source:QualifiedName "->"
  targets:QualifiedName ("," targets:QualifiedName)* ";";

/**
* A simple way to connect ports.
*
* @attribute source the source port or component instance
*   name
* @attribute targets a list of target port or component
*   instance names
*/
/ArcSimpleConnector =
  source:QualifiedName "->" targets:QualifiedName
                        ("," targets:QualifiedName)*;

/**
* ArcParameters are used in configurable components.
*
* @attribute Type the type of the parameter
* @attribute name the name of the parameter
*/
/ArcParameter =
  Type Name;

/**
* ArcConfigurationParameter are used to configure
* configurable components. It is either a value or a
* variable name.
*
* @attribute typeRef reference to an Enum type or
*     static constant
* @attribute variable a variable name
* @attribute value a concrete literal value
*/
/ArcConfigurationParameter =
  (Name ".")=> typeRef:QualifiedName |
  variable:Name |
  value:Literal;

/* =======================================================*/
/* ================== ASTRULES ========================*/
/* =======================================================*/
/ / replacement of ASTCNode with UMLPNode
ast ArcComponent astextends
    /mc.umlp.common._ast.UMLPNode;
ast ArcComponentHead astextends
```

```
216        /mc.umlp.common._ast.UMLPNode;
217   ast ArcComponentBody astextends
218        /mc.umlp.common._ast.UMLPNode;
219   ast ArcPort astextends
220        /mc.umlp.common._ast.UMLPNode;
221   ast ArcConnector astextends
222        /mc.umlp.common._ast.UMLPNode;
223   ast ArcSimpleConnector astextends
224        /mc.umlp.common._ast.UMLPNode;
225   ast ArcSubComponent astextends
226        /mc.umlp.common._ast.UMLPNode;
227   ast ArcSubComponentInstance astextends
228        /mc.umlp.common._ast.UMLPNode;
229   ast ArcParameter astextends
230        /mc.umlp.common._ast.UMLPNode;
231 }
```

Listing B.1: Common MontiCore grammar for architectural diagrams

## B.2 MontiArc Grammar

Grammar for MontiArc

```
package mc.umlp.arc;

version "1.0";
grammar MontiArc extends mc.umlp.arcd.ArchitectureDiagram {

/* =====================================================*/
/* ===================== OPTIONS =====================*/
/* =====================================================*/
  options {
    compilationunit ArcComponent
    nostring
    parser lookahead=5
    lexer lookahead=7
  }

/* =====================================================*/
/* ==================== PRODUCTIONS ==================*/
/* =====================================================*/

  /**
  * MontiArc components may contain arbitrary many
  * configurations. These configurations have to
  * implement this interface.
  */
  interface MontiArcConfig;

  /**
  * Extends the component body from super grammar with a
  * configuration.
  *
  * @attribute MontiArcConfig configures the component
  * @attribute ArcElement the architectural elements in the
  *   body
  * @Overwrite ArchitectureDiagram.ArcComponentBody
  */
  ArcComponentBody =
    "{"
      MontiArcConfig*
      ArcElement*
    "}";

  /**
  * An invariant constrains the behavior of a component.
  *
  * @attribute kind the optional kind of this invariant.
```

```
 * @attribute name name of the invariant
 * @attribute invariantExpression the invariant defined
 *   in the language 'kind'
 */
MontiArcInvariant implements
     (Name? "inv" Name ":")=> ArcElement =
  (kind:Name)? "inv" Name ":"
  invariantExpression:InvariantContent(parameter kind)
     ";";

/**
 * AutoConnect is used to connect ports automatically.
 *
 * @attribute stereotype optional stereotype
 * @attribute type autoconnect unambigous ports with the
 *   same type
 * @attribute port autoconnect unambigous ports with the
 *   same name and compatible type
 * @attribute off do not use autoconnection (default)
 */
MontiArcAutoConnect implements MontiArcConfig =
  "autoconnect" Stereotype?
  (["type"] | ["port"] | ["off"]) ";";

/**
 * Autoinstantiate is used to instantiate inner components
 * without generic parameters or configuration parameters
 *  automatically. If more then one instance of this inner
 *  component is created by using a reference, the
 * automatically instanciated reference will dissapear.
 *
 * @attribute stereotype optional stereotype
 * @attribute on turns autoinstantiate on
 * @attribute off turns autoinstantiate off (default)
 */
MontiArcAutoInstantiate implements MontiArcConfig  =
  "autoinstantiate" Stereotype?
  (["on"] | ["off"]) ";";

/**
 * Sets the time behaviour from the component.
 *
 * @attribute stereotype optional stereotype
 * @attribute timed a timed component
 * @attribute untimed an untimed component
 * @attribute timesynchronous a timesynchronous component
 *   (can only process one message per timeunit)
 */
MontiArcTimingParadigm implements MontiArcConfig  =
```

```
94      "behavior" Stereotype?
95      (["timed"] | ["untimed"] | ["timesynchronous"]) ";";
96
97 /* =======================================================*/
98 /* ================== ASTRULES ========================*/
99 /* =======================================================*/
100   // toString for ArcInvariant
101   ast MontiArcInvariant astextends
102       /mc.umlp.common._ast.UMLPNode =
103     method public String toString() {
104       return (this.getKind() != null ?
105           this.getKind() + " " : "") + "inv " +
106           this.getName();
107       };
108   // replacement of ASTCNode with UMLPNode
109   ast MontiArcTimingParadigm astextends
110       /mc.umlp.common._ast.UMLPNode;
111   ast MontiArcAutoInstantiate astextends
112       /mc.umlp.common._ast.UMLPNode;
113   ast MontiArcAutoConnect astextends
114       /mc.umlp.common._ast.UMLPNode;
115 }
```

Listing B.2: Common MontiCore grammar for MontiArc

# List of Figures

# Listings

## Aachener Informatik-Berichte

**This is the list of all technical reports since 1987. To obtain copies of reports please consult**

**http://aib.informatik.rwth-aachen.de/**

**or send your request to:**

**Informatik-Bibliothek, RWTH Aachen, Ahornstr. 55, 52056 Aachen,**
**Email:** biblio@informatik.rwth-aachen.de

1987-01 * Fachgruppe Informatik: Jahresbericht 1986
1987-02 * David de Frutos Escrig, Klaus Indermark: Equivalence Relations of Non-Deterministic Ianov-Schemes
1987-03 * Manfred Nagl: A Software Development Environment based on Graph Technology
1987-04 * Claus Lewerentz, Manfred Nagl, Bernhard Westfechtel: On Integration Mechanisms within a Graph-Based Software Development Environment
1987-05 * Reinhard Rinn: Über Eingabeanomalien bei verschiedenen Inferenzmodellen
1987-06 * Werner Damm, Gert Döhmen: Specifying Distributed Computer Architectures in AADL*
1987-07 * Gregor Engels, Claus Lewerentz, Wilhelm Schäfer: Graph Grammar Engineering: A Software Specification Method
1987-08 * Manfred Nagl: Set Theoretic Approaches to Graph Grammars
1987-09 * Claus Lewerentz, Andreas Schürr: Experiences with a Database System for Software Documents
1987-10 * Herbert Klaeren, Klaus Indermark: A New Implementation Technique for Recursive Function Definitions
1987-11 * Rita Loogen: Design of a Parallel Programmable Graph Reduction Machine with Distributed Memory
1987-12 J. Börstler, U. Möncke, R. Wilhelm: Table compression for tree automata
1988-01 * Gabriele Esser, Johannes Rückert, Frank Wagner Gesellschaftliche Aspekte der Informatik
1988-02 * Peter Martini, Otto Spaniol: Token-Passing in High-Speed Backbone Networks for Campus-Wide Environments
1988-03 * Thomas Welzel: Simulation of a Multiple Token Ring Backbone
1988-04 * Peter Martini: Performance Comparison for HSLAN Media Access Protocols
1988-05 * Peter Martini: Performance Analysis of Multiple Token Rings
1988-06 * Andreas Mann, Johannes Rückert, Otto Spaniol: Datenfunknetze
1988-07 * Andreas Mann, Johannes Rückert: Packet Radio Networks for Data Exchange
1988-08 * Andreas Mann, Johannes Rückert: Concurrent Slot Assignment Protocol for Packet Radio Networks
1988-09 * W. Kremer, F. Reichert, J. Rückert, A. Mann: Entwurf einer Netzwerktopologie für ein Mobilfunknetz zur Unterstützung des öffentlichen Straßenverkehrs

1988-10 * Kai Jakobs: Towards User-Friendly Networking
1988-11 * Kai Jakobs: The Directory - Evolution of a Standard
1988-12 * Kai Jakobs: Directory Services in Distributed Systems - A Survey
1988-13 * Martine Schümmer: RS-511, a Protocol for the Plant Floor
1988-14 * U. Quernheim: Satellite Communication Protocols - A Performance Comparison Considering On-Board Processing
1988-15 * Peter Martini, Otto Spaniol, Thomas Welzel: File Transfer in High Speed Token Ring Networks: Performance Evaluation by Approximate Analysis and Simulation
1988-16 * Fachgruppe Informatik: Jahresbericht 1987
1988-17 * Wolfgang Thomas: Automata on Infinite Objects
1988-18 * Michael Sonnenschein: On Petri Nets and Data Flow Graphs
1988-19 * Heiko Vogler: Functional Distribution of the Contextual Analysis in Block-Structured Programming Languages: A Case Study of Tree Transducers
1988-20 * Thomas Welzel: Einsatz des Simulationswerkzeuges QNAP2 zur Leistungsbewertung von Kommunikationsprotokollen
1988-21 * Th. Janning, C. Lewerentz: Integrated Project Team Management in a Software Development Environment
1988-22 * Joost Engelfriet, Heiko Vogler: Modular Tree Transducers
1988-23 * Wolfgang Thomas: Automata and Quantifier Hierarchies
1988-24 * Uschi Heuter: Generalized Definite Tree Languages
1989-01 * Fachgruppe Informatik: Jahresbericht 1988
1989-02 * G. Esser, J. Rückert, F. Wagner (Hrsg.): Gesellschaftliche Aspekte der Informatik
1989-03 * Heiko Vogler: Bottom-Up Computation of Primitive Recursive Tree Functions
1989-04 * Andy Schürr: Introduction to PROGRESS, an Attribute Graph Grammar Based Specification Language
1989-05 J. Börstler: Reuse and Software Development - Problems, Solutions, and Bibliography (in German)
1989-06 * Kai Jakobs: OSI - An Appropriate Basis for Group Communication?
1989-07 * Kai Jakobs: ISO's Directory Proposal - Evolution, Current Status and Future Problems
1989-08 * Bernhard Westfechtel: Extension of a Graph Storage for Software Documents with Primitives for Undo/Redo and Revision Control
1989-09 * Peter Martini: High Speed Local Area Networks - A Tutorial
1989-10 * P. Davids, Th. Welzel: Performance Analysis of DQDB Based on Simulation
1989-11 * Manfred Nagl (Ed.): Abstracts of Talks presented at the WG '89 15th International Workshop on Graphtheoretic Concepts in Computer Science
1989-12 * Peter Martini: The DQDB Protocol - Is it Playing the Game?
1989-13 * Martine Schümmer: CNC/DNC Communication with MAP
1989-14 * Martine Schümmer: Local Area Networks for Manufactoring Environments with hard Real-Time Requirements

1989-15 * M. Schümmer, Th. Welzel, P. Martini: Integration of Field Bus and MAP Networks - Hierarchical Communication Systems in Production Environments

1989-16 * G. Vossen, K.-U. Witt: SUXESS: Towards a Sound Unification of Extensions of the Relational Data Model

1989-17 * J. Derissen, P. Hruschka, M.v.d. Beeck, Th. Janning, M. Nagl: Integrating Structured Analysis and Information Modelling

1989-18 A. Maassen: Programming with Higher Order Functions

1989-19 * Mario Rodriguez-Artalejo, Heiko Vogler: A Narrowing Machine for Syntax Directed BABEL

1989-20 H. Kuchen, R. Loogen, J.J. Moreno Navarro, M. Rodriguez Artalejo: Graph-based Implementation of a Functional Logic Language

1990-01 * Fachgruppe Informatik: Jahresbericht 1989

1990-02 * Vera Jansen, Andreas Potthoff, Wolfgang Thomas, Udo Wermuth: A Short Guide to the AMORE System (Computing Automata, MOnoids and Regular Expressions)

1990-03 * Jerzy Skurczynski: On Three Hierarchies of Weak SkS Formulas

1990-04 R. Loogen: Stack-based Implementation of Narrowing

1990-05 H. Kuchen, A. Wagener: Comparison of Dynamic Load Balancing Strategies

1990-06 * Kai Jakobs, Frank Reichert: Directory Services for Mobile Communication

1990-07 * Kai Jakobs: What's Beyond the Interface - OSI Networks to Support Cooperative Work

1990-08 * Kai Jakobs: Directory Names and Schema - An Evaluation

1990-09 * Ulrich Quernheim, Dieter Kreuer: Das CCITT - Signalisierungssystem Nr. 7 auf Satellitenstrecken; Simulation der Zeichengabestrecke

1990-11 H. Kuchen, R. Loogen, J.J. Moreno Navarro, M. Rodriguez Artalejo: Lazy Narrowing in a Graph Machine

1990-12 * Kai Jakobs, Josef Kaltwasser, Frank Reichert, Otto Spaniol: Der Computer fährt mit

1990-13 * Rudolf Mathar, Andreas Mann: Analyzing a Distributed Slot Assignment Protocol by Markov Chains

1990-14 A. Maassen: Compilerentwicklung in Miranda - ein Praktikum in funktionaler Programmierung (written in german)

1990-15 * Manfred Nagl, Andreas Schürr: A Specification Environment for Graph Grammars

1990-16 A. Schürr: PROGRESS: A VHL-Language Based on Graph Grammars

1990-17 * Marita Möller: Ein Ebenenmodell wissensbasierter Konsultationen - Unterstützung für Wissensakquisition und Erklärungsfähigkeit

1990-18 * Eric Kowalewski: Entwurf und Interpretation einer Sprache zur Beschreibung von Konsultationsphasen in Expertensystemen

1990-20 Y. Ortega Mallen, D. de Frutos Escrig: A Complete Proof System for Timed Observations

1990-21 * Manfred Nagl: Modelling of Software Architectures: Importance, Notions, Experiences

1990-22 H. Fassbender, H. Vogler: A Call-by-need Implementation of Syntax Directed Functional Programming

1991-01 Guenther Geiler (ed.), Fachgruppe Informatik: Jahresbericht 1990
1991-03 B. Steffen, A. Ingolfsdottir: Characteristic Formulae for Processes with Divergence
1991-04 M. Portz: A new class of cryptosystems based on interconnection networks
1991-05 H. Kuchen, G. Geiler: Distributed Applicative Arrays
1991-06 * Ludwig Staiger: Kolmogorov Complexity and Hausdorff Dimension
1991-07 * Ludwig Staiger: Syntactic Congruences for w-languages
1991-09 * Eila Kuikka: A Proposal for a Syntax-Directed Text Processing System
1991-10 K. Gladitz, H. Fassbender, H. Vogler: Compiler-based Implementation of Syntax-Directed Functional Programming
1991-11 R. Loogen, St. Winkler: Dynamic Detection of Determinism in Functional Logic Languages
1991-12 * K. Indermark, M. Rodriguez Artalejo (Eds.): Granada Workshop on the Integration of Functional and Logic Programming
1991-13 * Rolf Hager, Wolfgang Kremer: The Adaptive Priority Scheduler: A More Fair Priority Service Discipline
1991-14 * Andreas Fasbender, Wolfgang Kremer: A New Approximation Algorithm for Tandem Networks with Priority Nodes
1991-15 J. Börstler, A. Zündorf: Revisiting extensions to Modula-2 to support reusability
1991-16 J. Börstler, Th. Janning: Bridging the gap between Requirements Analysis and Design
1991-17 A. Zündorf, A. Schürr: Nondeterministic Control Structures for Graph Rewriting Systems
1991-18 * Matthias Jarke, John Mylopoulos, Joachim W. Schmidt, Yannis Vassiliou: DAIDA: An Environment for Evolving Information Systems
1991-19 M. Jeusfeld, M. Jarke: From Relational to Object-Oriented Integrity Simplification
1991-20 G. Hogen, A. Kindler, R. Loogen: Automatic Parallelization of Lazy Functional Programs
1991-21 * Prof. Dr. rer. nat. Otto Spaniol: ODP (Open Distributed Processing): Yet another Viewpoint
1991-22 H. Kuchen, F. Lücking, H. Stoltze: The Topology Description Language TDL
1991-23 S. Graf, B. Steffen: Compositional Minimization of Finite State Systems
1991-24 R. Cleaveland, J. Parrow, B. Steffen: The Concurrency Workbench: A Semantics Based Tool for the Verification of Concurrent Systems
1991-25 * Rudolf Mathar, Jürgen Mattfeldt: Optimal Transmission Ranges for Mobile Communication in Linear Multihop Packet Radio Networks
1991-26 M. Jeusfeld, M. Staudt: Query Optimization in Deductive Object Bases
1991-27 J. Knoop, B. Steffen: The Interprocedural Coincidence Theorem
1991-28 J. Knoop, B. Steffen: Unifying Strength Reduction and Semantic Code Motion
1991-30 T. Margaria: First-Order theories for the verification of complex FSMs
1991-31 B. Steffen: Generating Data Flow Analysis Algorithms from Modal Specifications
1992-01 Stefan Eherer (ed.), Fachgruppe Informatik: Jahresbericht 1991

1992-02 * Bernhard Westfechtel: Basismechanismen zur Datenverwaltung in strukturbezogenen Hypertextsystemen

1992-04 S. A. Smolka, B. Steffen: Priority as Extremal Probability

1992-05 * Matthias Jarke, Carlos Maltzahn, Thomas Rose: Sharing Processes: Team Coordination in Design Repositories

1992-06 O. Burkart, B. Steffen: Model Checking for Context-Free Processes

1992-07 * Matthias Jarke, Klaus Pohl: Information Systems Quality and Quality Information Systems

1992-08 * Rudolf Mathar, Jürgen Mattfeldt: Analyzing Routing Strategy NFP in Multihop Packet Radio Networks on a Line

1992-09 * Alfons Kemper, Guido Moerkotte: Grundlagen objektorientierter Datenbanksysteme

1992-10 Matthias Jarke, Manfred Jeusfeld, Andreas Miethsam, Michael Gocek: Towards a logic-based reconstruction of software configuration management

1992-11 Werner Hans: A Complete Indexing Scheme for WAM-based Abstract Machines

1992-12 W. Hans, R. Loogen, St. Winkler: On the Interaction of Lazy Evaluation and Backtracking

1992-13 * Matthias Jarke, Thomas Rose: Specification Management with CAD

1992-14 Th. Noll, H. Vogler: Top-down Parsing with Simultaneous Evaluation on Noncircular Attribute Grammars

1992-15 A. Schuerr, B. Westfechtel: Graphgrammatiken und Graphersetzungssysteme(written in german)

1992-16 * Graduiertenkolleg Informatik und Technik (Hrsg.): Forschungsprojekte des Graduiertenkollegs Informatik und Technik

1992-17 M. Jarke (ed.): ConceptBase V3.1 User Manual

1992-18 * Clarence A. Ellis, Matthias Jarke (Eds.): Distributed Cooperation in Integrated Information Systems - Proceedings of the Third International Workshop on Intelligent and Cooperative Information Systems

1992-19-00 H. Kuchen, R. Loogen (eds.): Proceedings of the 4th Int. Workshop on the Parallel Implementation of Functional Languages

1992-19-01 G. Hogen, R. Loogen: PASTEL - A Parallel Stack-Based Implementation of Eager Functional Programs with Lazy Data Structures (Extended Abstract)

1992-19-02 H. Kuchen, K. Gladitz: Implementing Bags on a Shared Memory MIMD-Machine

1992-19-03 C. Rathsack, S.B. Scholz: LISA - A Lazy Interpreter for a Full-Fledged Lambda-Calculus

1992-19-04 T.A. Bratvold: Determining Useful Parallelism in Higher Order Functions

1992-19-05 S. Kahrs: Polymorphic Type Checking by Interpretation of Code

1992-19-06 M. Chakravarty, M. Köhler: Equational Constraints, Residuation, and the Parallel JUMP-Machine

1992-19-07 J. Seward: Polymorphic Strictness Analysis using Frontiers (Draft Version)

1992-19-08 D. Gärtner, A. Kimms, W. Kluge: pi-Redˆ+ - A Compiling Graph-Reduction System for a Full Fledged Lambda-Calculus

1992-19-09 D. Howe, G. Burn: Experiments with strict STG code

1992-19-10 J. Glauert: Parallel Implementation of Functional Languages Using Small Processes

1992-19-11 M. Joy, T. Axford: A Parallel Graph Reduction Machine

1992-19-12 A. Bennett, P. Kelly: Simulation of Multicache Parallel Reduction

1992-19-13 K. Langendoen, D.J. Agterkamp: Cache Behaviour of Lazy Functional Programs (Working Paper)

1992-19-14 K. Hammond, S. Peyton Jones: Profiling scheduling strategies on the GRIP parallel reducer

1992-19-15 S. Mintchev: Using Strictness Information in the STG-machine

1992-19-16 D. Rushall: An Attribute Grammar Evaluator in Haskell

1992-19-17 J. Wild, H. Glaser, P. Hartel: Statistics on storage management in a lazy functional language implementation

1992-19-18 W.S. Martins: Parallel Implementations of Functional Languages

1992-19-19 D. Lester: Distributed Garbage Collection of Cyclic Structures (Draft version)

1992-19-20 J.C. Glas, R.F.H. Hofman, W.G. Vree: Parallelization of Branch-and-Bound Algorithms in a Functional Programming Environment

1992-19-21 S. Hwang, D. Rushall: The nu-STG machine: a parallelized Spineless Tagless Graph Reduction Machine in a distributed memory architecture (Draft version)

1992-19-22 G. Burn, D. Le Metayer: Cps-Translation and the Correctness of Optimising Compilers

1992-19-23 S.L. Peyton Jones, P. Wadler: Imperative functional programming (Brief summary)

1992-19-24 W. Damm, F. Liu, Th. Peikenkamp: Evaluation and Parallelization of Functions in Functional + Logic Languages (abstract)

1992-19-25 M. Kesseler: Communication Issues Regarding Parallel Functional Graph Rewriting

1992-19-26 Th. Peikenkamp: Charakterizing and representing neededness in functional loginc languages (abstract)

1992-19-27 H. Doerr: Monitoring with Graph-Grammars as formal operational Models

1992-19-28 J. van Groningen: Some implementation aspects of Concurrent Clean on distributed memory architectures

1992-19-29 G. Ostheimer: Load Bounding for Implicit Parallelism (abstract)

1992-20 H. Kuchen, F.J. Lopez Fraguas, J.J. Moreno Navarro, M. Rodriguez Artalejo: Implementing Disequality in a Lazy Functional Logic Language

1992-21 H. Kuchen, F.J. Lopez Fraguas: Result Directed Computing in a Functional Logic Language

1992-22 H. Kuchen, J.J. Moreno Navarro, M.V. Hermenegildo: Independent AND-Parallel Narrowing

1992-23 T. Margaria, B. Steffen: Distinguishing Formulas for Free

1992-24 K. Pohl: The Three Dimensions of Requirements Engineering

1992-25 * R. Stainov: A Dynamic Configuration Facility for Multimedia Communications

1992-26 * Michael von der Beeck: Integration of Structured Analysis and Timed Statecharts for Real-Time and Concurrency Specification

1992-27 W. Hans, St. Winkler: Aliasing and Groundness Analysis of Logic Programs through Abstract Interpretation and its Safety

1992-28 * Gerhard Steinke, Matthias Jarke: Support for Security Modeling in Information Systems Design

1992-29 B. Schinzel: Warum Frauenforschung in Naturwissenschaft und Technik

1992-30 A. Kemper, G. Moerkotte, K. Peithner: Object-Orientation Axiomatised by Dynamic Logic

1992-32 * Bernd Heinrichs, Kai Jakobs: Timer Handling in High-Performance Transport Systems

1992-33 * B. Heinrichs, K. Jakobs, K. Lenßen, W. Reinhardt, A. Spinner: Euro-Bridge: Communication Services for Multimedia Applications

1992-34 C. Gerlhof, A. Kemper, Ch. Kilger, G. Moerkotte: Partition-Based Clustering in Object Bases: From Theory to Practice

1992-35 J. Börstler: Feature-Oriented Classification and Reuse in IPSEN

1992-36 M. Jarke, J. Bubenko, C. Rolland, A. Sutcliffe, Y. Vassiliou: Theories Underlying Requirements Engineering: An Overview of NATURE at Genesis

1992-37 * K. Pohl, M. Jarke: Quality Information Systems: Repository Support for Evolving Process Models

1992-38 A. Zuendorf: Implementation of the imperative / rule based language PROGRES

1992-39 P. Koch: Intelligentes Backtracking bei der Auswertung funktional-logischer Programme

1992-40 * Rudolf Mathar, Jürgen Mattfeldt: Channel Assignment in Cellular Radio Networks

1992-41 * Gerhard Friedrich, Wolfgang Neidl: Constructive Utility in Model-Based Diagnosis Repair Systems

1992-42 * P. S. Chen, R. Hennicker, M. Jarke: On the Retrieval of Reusable Software Components

1992-43 W. Hans, St.Winkler: Abstract Interpretation of Functional Logic Languages

1992-44 N. Kiesel, A. Schuerr, B. Westfechtel: Design and Evaluation of GRAS, a Graph-Oriented Database System for Engineering Applications

1993-01 * Fachgruppe Informatik: Jahresbericht 1992

1993-02 * Patrick Shicheng Chen: On Inference Rules of Logic-Based Information Retrieval Systems

1993-03 G. Hogen, R. Loogen: A New Stack Technique for the Management of Runtime Structures in Distributed Environments

1993-05 A. Zündorf: A Heuristic for the Subgraph Isomorphism Problem in Executing PROGRES

1993-06 A. Kemper, D. Kossmann: Adaptable Pointer Swizzling Strategies in Object Bases: Design, Realization, and Quantitative Analysis

1993-07 * Graduiertenkolleg Informatik und Technik (Hrsg.): Graduiertenkolleg Informatik und Technik

1993-08 * Matthias Berger: k-Coloring Vertices using a Neural Network with Convergence to Valid Solutions

1993-09 M. Buchheit, M. Jeusfeld, W. Nutt, M. Staudt: Subsumption between Queries to Object-Oriented Databases

1993-10 O. Burkart, B. Steffen: Pushdown Processes: Parallel Composition and Model Checking

1993-11 * R. Große-Wienker, O. Hermanns, D. Menzenbach, A. Pollacks, S. Repetzki, J. Schwartz, K. Sonnenschein, B. Westfechtel: Das SUKITS-Projekt: A-posteriori-Integration heterogener CIM-Anwendungssysteme

1993-12 * Rudolf Mathar, Jürgen Mattfeldt: On the Distribution of Cumulated Interference Power in Rayleigh Fading Channels

1993-13 O. Maler, L. Staiger: On Syntactic Congruences for omega-languages

1993-14 M. Jarke, St. Eherer, R. Gallersdoerfer, M. Jeusfeld, M. Staudt: ConceptBase - A Deductive Object Base Manager

1993-15 M. Staudt, H.W. Nissen, M.A. Jeusfeld: Query by Class, Rule and Concept

1993-16 * M. Jarke, K. Pohl, St. Jacobs et al.: Requirements Engineering: An Integrated View of Representation Process and Domain

1993-17 * M. Jarke, K. Pohl: Establishing Vision in Context: Towards a Model of Requirements Processes

1993-18 W. Hans, H. Kuchen, St. Winkler: Full Indexing for Lazy Narrowing

1993-19 W. Hans, J.J. Ruz, F. Saenz, St. Winkler: A VHDL Specification of a Shared Memory Parallel Machine for Babel

1993-20 * K. Finke, M. Jarke, P. Szczurko, R. Soltysiak: Quality Management for Expert Systems in Process Control

1993-21 M. Jarke, M.A. Jeusfeld, P. Szczurko: Three Aspects of Intelligent Cooperation in the Quality Cycle

1994-01 Margit Generet, Sven Martin (eds.), Fachgruppe Informatik: Jahresbericht 1993

1994-02 M. Lefering: Development of Incremental Integration Tools Using Formal Specifications

1994-03 * P. Constantopoulos, M. Jarke, J. Mylopoulos, Y. Vassiliou: The Software Information Base: A Server for Reuse

1994-04 * Rolf Hager, Rudolf Mathar, Jürgen Mattfeldt: Intelligent Cruise Control and Reliable Communication of Mobile Stations

1994-05 * Rolf Hager, Peter Hermesmann, Michael Portz: Feasibility of Authentication Procedures within Advanced Transport Telematics

1994-06 * Claudia Popien, Bernd Meyer, Axel Kuepper: A Formal Approach to Service Import in ODP Trader Federations

1994-07 P. Peters, P. Szczurko: Integrating Models of Quality Management Methods by an Object-Oriented Repository

1994-08 * Manfred Nagl, Bernhard Westfechtel: A Universal Component for the Administration in Distributed and Integrated Development Environments

1994-09 * Patrick Horster, Holger Petersen: Signatur- und Authentifikationsverfahren auf der Basis des diskreten Logarithmusproblems

1994-11 A. Schürr: PROGRES, A Visual Language and Environment for PROgramming with Graph REwrite Systems

1994-12 A. Schürr: Specification of Graph Translators with Triple Graph Grammars

1994-13 A. Schürr: Logic Based Programmed Structure Rewriting Systems

1994-14 L. Staiger: Codes, Simplifying Words, and Open Set Condition

1994-15 * Bernhard Westfechtel: A Graph-Based System for Managing Configurations of Engineering Design Documents
1994-16 P. Klein: Designing Software with Modula-3
1994-17 I. Litovsky, L. Staiger: Finite acceptance of infinite words
1994-18 G. Hogen, R. Loogen: Parallel Functional Implementations: Graphbased vs. Stackbased Reduction
1994-19 M. Jeusfeld, U. Johnen: An Executable Meta Model for Re-Engineering of Database Schemas
1994-20 * R. Gallersdörfer, M. Jarke, K. Klabunde: Intelligent Networks as a Data Intensive Application (INDIA)
1994-21 M. Mohnen: Proving the Correctness of the Static Link Technique Using Evolving Algebras
1994-22 H. Fernau, L. Staiger: Valuations and Unambiguity of Languages, with Applications to Fractal Geometry
1994-24 * M. Jarke, K. Pohl, R. Dömges, St. Jacobs, H. W. Nissen: Requirements Information Management: The NATURE Approach
1994-25 * M. Jarke, K. Pohl, C. Rolland, J.-R. Schmitt: Experience-Based Method Evaluation and Improvement: A Process Modeling Approach
1994-26 * St. Jacobs, St. Kethers: Improving Communication and Decision Making within Quality Function Deployment
1994-27 * M. Jarke, H. W. Nissen, K. Pohl: Tool Integration in Evolving Information Systems Environments
1994-28 O. Burkart, D. Caucal, B. Steffen: An Elementary Bisimulation Decision Procedure for Arbitrary Context-Free Processes
1995-01 * Fachgruppe Informatik: Jahresbericht 1994
1995-02 Andy Schürr, Andreas J. Winter, Albert Zündorf: Graph Grammar Engineering with PROGRES
1995-03 Ludwig Staiger: A Tight Upper Bound on Kolmogorov Complexity by Hausdorff Dimension and Uniformly Optimal Prediction
1995-04 Birgitta König-Ries, Sven Helmer, Guido Moerkotte: An experimental study on the complexity of left-deep join ordering problems for cyclic queries
1995-05 Sophie Cluet, Guido Moerkotte: Efficient Evaluation of Aggregates on Bulk Types
1995-06 Sophie Cluet, Guido Moerkotte: Nested Queries in Object Bases
1995-07 Sophie Cluet, Guido Moerkotte: Query Optimization Techniques Exploiting Class Hierarchies
1995-08 Markus Mohnen: Efficient Compile-Time Garbage Collection for Arbitrary Data Structures
1995-09 Markus Mohnen: Functional Specification of Imperative Programs: An Alternative Point of View of Functional Languages
1995-10 Rainer Gallersdörfer, Matthias Nicola: Improving Performance in Replicated Databases through Relaxed Coherency
1995-11 * M.Staudt, K.von Thadden: Subsumption Checking in Knowledge Bases
1995-12 * G.V.Zemanek, H.W.Nissen, H.Hubert, M.Jarke: Requirements Analysis from Multiple Perspectives: Experiences with Conceptual Modeling Technology

1995-13 * M.Staudt, M.Jarke: Incremental Maintenance of Externally Materialized Views
1995-14 * P.Peters, P.Szczurko, M.Jeusfeld: Oriented Information Management: Conceptual Models at Work
1995-15 * Matthias Jarke, Sudha Ram (Hrsg.): WITS 95 Proceedings of the 5th Annual Workshop on Information Technologies and Systems
1995-16 * W.Hans, St.Winkler, F.Saenz: Distributed Execution in Functional Logic Programming
1996-01 * Jahresbericht 1995
1996-02 Michael Hanus, Christian Prehofer: Higher-Order Narrowing with Definitional Trees
1996-03 * W.Scheufele, G.Moerkotte: Optimal Ordering of Selections and Joins in Acyclic Queries with Expensive Predicates
1996-04 Klaus Pohl: PRO-ART: Enabling Requirements Pre-Traceability
1996-05 Klaus Pohl: Requirements Engineering: An Overview
1996-06 * M.Jarke, W.Marquardt: Design and Evaluation of Computer–Aided Process Modelling Tools
1996-07 Olaf Chitil: The Sigma-Semantics: A Comprehensive Semantics for Functional Programs
1996-08 * S.Sripada: On Entropy and the Limitations of the Second Law of Thermodynamics
1996-09 Michael Hanus (Ed.): Proceedings of the Poster Session of ALP96 - Fifth International Conference on Algebraic and Logic Programming
1996-09-0 Michael Hanus (Ed.): Proceedings of the Poster Session of ALP 96 - Fifth International Conference on Algebraic and Logic Programming: Introduction and table of contents
1996-09-1 Ilies Alouini: An Implementation of Conditional Concurrent Rewriting on Distributed Memory Machines
1996-09-2 Olivier Danvy, Karoline Malmkjær: On the Idempotence of the CPS Transformation
1996-09-3 Víctor M. Gulías, José L. Freire: Concurrent Programming in Haskell
1996-09-4 Sébastien Limet, Pierre Réty: On Decidability of Unifiability Modulo Rewrite Systems
1996-09-5 Alexandre Tessier: Declarative Debugging in Constraint Logic Programming
1996-10 Reidar Conradi, Bernhard Westfechtel: Version Models for Software Configuration Management
1996-11 * C.Weise, D.Lenzkes: A Fast Decision Algorithm for Timed Refinement
1996-12 * R.Dömges, K.Pohl, M.Jarke, B.Lohmann, W.Marquardt: PRO-ART/CE* — An Environment for Managing the Evolution of Chemical Process Simulation Models
1996-13 * K.Pohl, R.Klamma, K.Weidenhaupt, R.Dömges, P.Haumer, M.Jarke: A Framework for Process-Integrated Tools
1996-14 * R.Gallersdörfer, K.Klabunde, A.Stolz, M.Eßmajor: INDIA — Intelligent Networks as a Data Intensive Application, Final Project Report, June 1996
1996-15 * H.Schimpe, M.Staudt: VAREX: An Environment for Validating and Refining Rule Bases

1996-16 * M.Jarke, M.Gebhardt, S.Jacobs, H.Nissen: Conflict Analysis Across Heterogeneous Viewpoints: Formalization and Visualization

1996-17 Manfred A. Jeusfeld, Tung X. Bui: Decision Support Components on the Internet

1996-18 Manfred A. Jeusfeld, Mike Papazoglou: Information Brokering: Design, Search and Transformation

1996-19 * P.Peters, M.Jarke: Simulating the impact of information flows in networked organizations

1996-20 Matthias Jarke, Peter Peters, Manfred A. Jeusfeld: Model-driven planning and design of cooperative information systems

1996-21 * G.de Michelis, E.Dubois, M.Jarke, F.Matthes, J.Mylopoulos, K.Pohl, J.Schmidt, C.Woo, E.Yu: Cooperative information systems: a manifesto

1996-22 * S.Jacobs, M.Gebhardt, S.Kethers, W.Rzasa: Filling HTML forms simultaneously: CoWeb architecture and functionality

1996-23 * M.Gebhardt, S.Jacobs: Conflict Management in Design

1997-01 Michael Hanus, Frank Zartmann (eds.): Jahresbericht 1996

1997-02 Johannes Faassen: Using full parallel Boltzmann Machines for Optimization

1997-03 Andreas Winter, Andy Schürr: Modules and Updatable Graph Views for PROgrammed Graph REwriting Systems

1997-04 Markus Mohnen, Stefan Tobies: Implementing Context Patterns in the Glasgow Haskell Compiler

1997-05 * S.Gruner: Schemakorrespondenzaxiome unterstützen die paargrammatische Spezifikation inkrementeller Integrationswerkzeuge

1997-06 Matthias Nicola, Matthias Jarke: Design and Evaluation of Wireless Health Care Information Systems in Developing Countries

1997-07 Petra Hofstedt: Taskparallele Skelette für irregulär strukturierte Probleme in deklarativen Sprachen

1997-08 Dorothea Blostein, Andy Schürr: Computing with Graphs and Graph Rewriting

1997-09 Carl-Arndt Krapp, Bernhard Westfechtel: Feedback Handling in Dynamic Task Nets

1997-10 Matthias Nicola, Matthias Jarke: Integrating Replication and Communication in Performance Models of Distributed Databases

1997-11 * R. Klamma, P. Peters, M. Jarke: Workflow Support for Failure Management in Federated Organizations

1997-13 Markus Mohnen: Optimising the Memory Management of Higher-Order Functional Programs

1997-14 Roland Baumann: Client/Server Distribution in a Structure-Oriented Database Management System

1997-15 George Botorog: High-Level Parallel Programming and the Efficient Implementation of Numerical Algorithms

1998-01 * Fachgruppe Informatik: Jahresbericht 1997

1998-02 Stefan Gruner, Manfred Nagel, Andy Schürr: Fine-grained and Structure-Oriented Document Integration Tools are Needed for Development Processes

| | | |
|---|---|---|
| 1998-03 | | Stefan Gruner: Einige Anmerkungen zur graphgrammatischen Spezifikation von Integrationswerkzeugen nach Westfechtel, Janning, Lefering und Schürr |
| 1998-04 | * | O. Kubitz: Mobile Robots in Dynamic Environments |
| 1998-05 | | Martin Leucker, Stephan Tobies: Truth - A Verification Platform for Distributed Systems |
| 1998-06 | * | Matthias Oliver Berger: DECT in the Factory of the Future |
| 1998-07 | | M. Arnold, M. Erdmann, M. Glinz, P. Haumer, R. Knoll, B. Paech, K. Pohl, J. Ryser, R. Studer, K. Weidenhaupt: Survey on the Scenario Use in Twelve Selected Industrial Projects |
| 1998-09 | * | Th. Lehmann: Geometrische Ausrichtung medizinischer Bilder am Beispiel intraoraler Radiographien |
| 1998-10 | * | M. Nicola, M. Jarke: Performance Modeling of Distributed and Replicated Databases |
| 1998-11 | * | Ansgar Schleicher, Bernhard Westfechtel, Dirk Jäger: Modeling Dynamic Software Processes in UML |
| 1998-12 | * | W. Appelt, M. Jarke: Interoperable Tools for Cooperation Support using the World Wide Web |
| 1998-13 | | Klaus Indermark: Semantik rekursiver Funktionsdefinitionen mit Striktheitsinformation |
| 1999-01 | * | Jahresbericht 1998 |
| 1999-02 | * | F. Huch: Verifcation of Erlang Programs using Abstract Interpretation and Model Checking — Extended Version |
| 1999-03 | * | R. Gallersdörfer, M. Jarke, M. Nicola: The ADR Replication Manager |
| 1999-04 | | María Alpuente, Michael Hanus, Salvador Lucas, Germán Vidal: Specialization of Functional Logic Programs Based on Needed Narrowing |
| 1999-05 | * | W. Thomas (Ed.): DLT 99 - Developments in Language Theory Fourth International Conference |
| 1999-06 | * | Kai Jakobs, Klaus-Dieter Kleefeld: Informationssysteme für die angewandte historische Geographie |
| 1999-07 | | Thomas Wilke: CTL+ is exponentially more succinct than CTL |
| 1999-08 | | Oliver Matz: Dot-Depth and Monadic Quantifier Alternation over Pictures |
| 2000-01 | * | Jahresbericht 1999 |
| 2000-02 | | Jens Vöge, Marcin Jurdzinski A Discrete Strategy Improvement Algorithm for Solving Parity Games |
| 2000-03 | | D. Jäger, A. Schleicher, B. Westfechtel: UPGRADE: A Framework for Building Graph-Based Software Engineering Tools |
| 2000-04 | | Andreas Becks, Stefan Sklorz, Matthias Jarke: Exploring the Semantic Structure of Technical Document Collections: A Cooperative Systems Approach |
| 2000-05 | | Mareike Schoop: Cooperative Document Management |
| 2000-06 | | Mareike Schoop, Christoph Quix (eds.): Proceedings of the Fifth International Workshop on the Language-Action Perspective on Communication Modelling |
| 2000-07 | * | Markus Mohnen, Pieter Koopman (Eds.): Proceedings of the 12th International Workshop of Functional Languages |

| | |
|---|---|
| 2000-08 | Thomas Arts, Thomas Noll: Verifying Generic Erlang Client-Server Implementations |
| 2001-01 * | Jahresbericht 2000 |
| 2001-02 | Benedikt Bollig, Martin Leucker: Deciding LTL over Mazurkiewicz Traces |
| 2001-03 | Thierry Cachat: The power of one-letter rational languages |
| 2001-04 | Benedikt Bollig, Martin Leucker, Michael Weber: Local Parallel Model Checking for the Alternation Free mu-Calculus |
| 2001-05 | Benedikt Bollig, Martin Leucker, Thomas Noll: Regular MSC Languages |
| 2001-06 | Achim Blumensath: Prefix-Recognisable Graphs and Monadic Second-Order Logic |
| 2001-07 | Martin Grohe, Stefan Wöhrle: An Existential Locality Theorem |
| 2001-08 | Mareike Schoop, James Taylor (eds.): Proceedings of the Sixth International Workshop on the Language-Action Perspective on Communication Modelling |
| 2001-09 | Thomas Arts, Jürgen Giesl: A collection of examples for termination of term rewriting using dependency pairs |
| 2001-10 | Achim Blumensath: Axiomatising Tree-interpretable Structures |
| 2001-11 | Klaus Indermark, Thomas Noll (eds.): Kolloquium Programmiersprachen und Grundlagen der Programmierung |
| 2002-01 * | Jahresbericht 2001 |
| 2002-02 | Jürgen Giesl, Aart Middeldorp: Transformation Techniques for Context-Sensitive Rewrite Systems |
| 2002-03 | Benedikt Bollig, Martin Leucker, Thomas Noll: Generalised Regular MSC Languages |
| 2002-04 | Jürgen Giesl, Aart Middeldorp: Innermost Termination of Context-Sensitive Rewriting |
| 2002-05 | Horst Lichter, Thomas von der Maßen, Thomas Weiler: Modelling Requirements and Architectures for Software Product Lines |
| 2002-06 | Henry N. Adorna: 3-Party Message Complexity is Better than 2-Party Ones for Proving Lower Bounds on the Size of Minimal Nondeterministic Finite Automata |
| 2002-07 | Jörg Dahmen: Invariant Image Object Recognition using Gaussian Mixture Densities |
| 2002-08 | Markus Mohnen: An Open Framework for Data-Flow Analysis in Java |
| 2002-09 | Markus Mohnen: Interfaces with Default Implementations in Java |
| 2002-10 | Martin Leucker: Logics for Mazurkiewicz traces |
| 2002-11 | Jürgen Giesl, Hans Zantema: Liveness in Rewriting |
| 2003-01 * | Jahresbericht 2002 |
| 2003-02 | Jürgen Giesl, René Thiemann: Size-Change Termination for Term Rewriting |
| 2003-03 | Jürgen Giesl, Deepak Kapur: Deciding Inductive Validity of Equations |
| 2003-04 | Jürgen Giesl, René Thiemann, Peter Schneider-Kamp, Stephan Falke: Improving Dependency Pairs |
| 2003-05 | Christof Löding, Philipp Rohde: Solving the Sabotage Game is PSPACE-hard |
| 2003-06 | Franz Josef Och: Statistical Machine Translation: From Single-Word Models to Alignment Templates |

2003-07 Horst Lichter, Thomas von der Maßen, Alexander Nyßen, Thomas Weiler: Vergleich von Ansätzen zur Feature Modellierung bei der Softwareproduktlinienentwicklung

2003-08 Jürgen Giesl, René Thiemann, Peter Schneider-Kamp, Stephan Falke: Mechanizing Dependency Pairs

2004-01 * Fachgruppe Informatik: Jahresbericht 2003

2004-02 Benedikt Bollig, Martin Leucker: Message-Passing Automata are expressively equivalent to EMSO logic

2004-03 Delia Kesner, Femke van Raamsdonk, Joe Wells (eds.): HOR 2004 – 2nd International Workshop on Higher-Order Rewriting

2004-04 Slim Abdennadher, Christophe Ringeissen (eds.): RULE 04 – Fifth International Workshop on Rule-Based Programming

2004-05 Herbert Kuchen (ed.): WFLP 04 – 13th International Workshop on Functional and (Constraint) Logic Programming

2004-06 Sergio Antoy, Yoshihito Toyama (eds.): WRS 04 – 4th International Workshop on Reduction Strategies in Rewriting and Programming

2004-07 Michael Codish, Aart Middeldorp (eds.): WST 04 – 7th International Workshop on Termination

2004-08 Klaus Indermark, Thomas Noll: Algebraic Correctness Proofs for Compiling Recursive Function Definitions with Strictness Information

2004-09 Joachim Kneis, Daniel Mölle, Stefan Richter, Peter Rossmanith: Parameterized Power Domination Complexity

2004-10 Zinaida Benenson, Felix C. Gärtner, Dogan Kesdogan: Secure Multi-Party Computation with Security Modules

2005-01 * Fachgruppe Informatik: Jahresbericht 2004

2005-02 Maximillian Dornseif, Felix C. Gärtner, Thorsten Holz, Martin Mink: An Offensive Approach to Teaching Information Security: "Aachen Summer School Applied IT Security"

2005-03 Jürgen Giesl, René Thiemann, Peter Schneider-Kamp: Proving and Disproving Termination of Higher-Order Functions

2005-04 Daniel Mölle, Stefan Richter, Peter Rossmanith: A Faster Algorithm for the Steiner Tree Problem

2005-05 Fabien Pouget, Thorsten Holz: A Pointillist Approach for Comparing Honeypots

2005-06 Simon Fischer, Berthold Vöcking: Adaptive Routing with Stale Information

2005-07 Felix C. Freiling, Thorsten Holz, Georg Wicherski: Botnet Tracking: Exploring a Root-Cause Methodology to Prevent Distributed Denial-of-Service Attacks

2005-08 Joachim Kneis, Peter Rossmanith: A New Satisfiability Algorithm With Applications To Max-Cut

2005-09 Klaus Kursawe, Felix C. Freiling: Byzantine Fault Tolerance on General Hybrid Adversary Structures

2005-10 Benedikt Bollig: Automata and Logics for Message Sequence Charts

2005-11 Simon Fischer, Berthold Vöcking: A Counterexample to the Fully Mixed Nash Equilibrium Conjecture

2005-12 Neeraj Mittal, Felix Freiling, S. Venkatesan, Lucia Draque Penso: Efficient Reductions for Wait-Free Termination Detection in Faulty Distributed Systems
2005-13 Carole Delporte-Gallet, Hugues Fauconnier, Felix C. Freiling: Revisiting Failure Detection and Consensus in Omission Failure Environments
2005-14 Felix C. Freiling, Sukumar Ghosh: Code Stabilization
2005-15 Uwe Naumann: The Complexity of Derivative Computation
2005-16 Uwe Naumann: Syntax-Directed Derivative Code (Part I: Tangent-Linear Code)
2005-17 Uwe Naumann: Syntax-directed Derivative Code (Part II: Intraprocedural Adjoint Code)
2005-18 Thomas von der Maßen, Klaus Müller, John MacGregor, Eva Geisberger, Jörg Dörr, Frank Houdek, Harbhajan Singh, Holger Wußmann, Hans-Veit Bacher, Barbara Paech: Einsatz von Features im Software-Entwicklungsprozess - Abschlußbericht des GI-Arbeitskreises “Features”
2005-19 Uwe Naumann, Andre Vehreschild: Tangent-Linear Code by Augmented LL-Parsers
2005-20 Felix C. Freiling, Martin Mink: Bericht über den Workshop zur Ausbildung im Bereich IT-Sicherheit Hochschulausbildung, berufliche Weiterbildung, Zertifizierung von Ausbildungsangeboten am 11. und 12. August 2005 in Köln organisiert von RWTH Aachen in Kooperation mit BITKOM, BSI, DLR und Gesellschaft fuer Informatik (GI) e.V.
2005-21 Thomas Noll, Stefan Rieger: Optimization of Straight-Line Code Revisited
2005-22 Felix Freiling, Maurice Herlihy, Lucia Draque Penso: Optimal Randomized Fair Exchange with Secret Shared Coins
2005-23 Heiner Ackermann, Alantha Newman, Heiko Röglin, Berthold Vöcking: Decision Making Based on Approximate and Smoothed Pareto Curves
2005-24 Alexander Becher, Zinaida Benenson, Maximillian Dornseif: Tampering with Motes: Real-World Physical Attacks on Wireless Sensor Networks
2006-01 * Fachgruppe Informatik: Jahresbericht 2005
2006-02 Michael Weber: Parallel Algorithms for Verification of Large Systems
2006-03 Michael Maier, Uwe Naumann: Intraprocedural Adjoint Code Generated by the Differentiation-Enabled NAGWare Fortran Compiler
2006-04 Ebadollah Varnik, Uwe Naumann, Andrew Lyons: Toward Low Static Memory Jacobian Accumulation
2006-05 Uwe Naumann, Jean Utke, Patrick Heimbach, Chris Hill, Derya Ozyurt, Carl Wunsch, Mike Fagan, Nathan Tallent, Michelle Strout: Adjoint Code by Source Transformation with OpenAD/F
2006-06 Joachim Kneis, Daniel Mölle, Stefan Richter, Peter Rossmanith: Divide-and-Color
2006-07 Thomas Colcombet, Christof Löding: Transforming structures by set interpretations
2006-08 Uwe Naumann, Yuxiao Hu: Optimal Vertex Elimination in Single-Expression-Use Graphs
2006-09 Tingting Han, Joost-Pieter Katoen: Counterexamples in Probabilistic Model Checking

2006-10 Mesut Günes, Alexander Zimmermann, Martin Wenig, Jan Ritzerfeld, Ulrich Meis: From Simulations to Testbeds - Architecture of the Hybrid MCG-Mesh Testbed

2006-11 Bastian Schlich, Michael Rohrbach, Michael Weber, Stefan Kowalewski: Model Checking Software for Microcontrollers

2006-12 Benedikt Bollig, Joost-Pieter Katoen, Carsten Kern, Martin Leucker: Replaying Play in and Play out: Synthesis of Design Models from Scenarios by Learning

2006-13 Wong Karianto, Christof Löding: Unranked Tree Automata with Sibling Equalities and Disequalities

2006-14 Danilo Beuche, Andreas Birk, Heinrich Dreier, Andreas Fleischmann, Heidi Galle, Gerald Heller, Dirk Janzen, Isabel John, Ramin Tavakoli Kolagari, Thomas von der Maßen, Andreas Wolfram: Report of the GI Work Group "Requirements Management Tools for Product Line Engineering"

2006-15 Sebastian Ullrich, Jakob T. Valvoda, Torsten Kuhlen: Utilizing optical sensors from mice for new input devices

2006-16 Rafael Ballagas, Jan Borchers: Selexels: a Conceptual Framework for Pointing Devices with Low Expressiveness

2006-17 Eric Lee, Henning Kiel, Jan Borchers: Scrolling Through Time: Improving Interfaces for Searching and Navigating Continuous Audio Timelines

2007-01 * Fachgruppe Informatik: Jahresbericht 2006

2007-02 Carsten Fuhs, Jürgen Giesl, Aart Middeldorp, Peter Schneider-Kamp, René Thiemann, and Harald Zankl: SAT Solving for Termination Analysis with Polynomial Interpretations

2007-03 Jürgen Giesl, René Thiemann, Stephan Swiderski, and Peter Schneider-Kamp: Proving Termination by Bounded Increase

2007-04 Jan Buchholz, Eric Lee, Jonathan Klein, and Jan Borchers: coJIVE: A System to Support Collaborative Jazz Improvisation

2007-05 Uwe Naumann: On Optimal DAG Reversal

2007-06 Joost-Pieter Katoen, Thomas Noll, and Stefan Rieger: Verifying Concurrent List-Manipulating Programs by LTL Model Checking

2007-07 Alexander Nyßen, Horst Lichter: MeDUSA - MethoD for UML2-based Design of Embedded Software Applications

2007-08 Falk Salewski and Stefan Kowalewski: Achieving Highly Reliable Embedded Software: An empirical evaluation of different approaches

2007-09 Tina Kraußer, Heiko Mantel, and Henning Sudbrock: A Probabilistic Justification of the Combining Calculus under the Uniform Scheduler Assumption

2007-10 Martin Neuhäußer, Joost-Pieter Katoen: Bisimulation and Logical Preservation for Continuous-Time Markov Decision Processes

2007-11 Klaus Wehrle (editor): 6. Fachgespräch Sensornetzwerke

2007-12 Uwe Naumann: An L-Attributed Grammar for Adjoint Code

2007-13 Uwe Naumann, Michael Maier, Jan Riehme, and Bruce Christianson: Second-Order Adjoints by Source Code Manipulation of Numerical Programs 2007-14 Jean Utke, Uwe Naumann, Mike Fagan, Nathan Tallent, Michelle Strout, Patrick Heimbach, Chris Hill, and Carl Wunsch: OpenAD/F: A Modular, Open-Source Tool for Automatic Differentiation of Fortran Codes
2007-15 Volker Stolz: Temporal assertions for sequential and concurrent programs
2007-16 Sadeq Ali Makram, Mesut Güneç, Martin Wenig, Alexander Zimmermann: Adaptive Channel Assignment to Support QoS and Load Balancing for Wireless Mesh Networks
2007-17 René Thiemann: The DP Framework for Proving Termination of Term Rewriting
2007-18 Uwe Naumann: Call Tree Reversal is NP-Complete
2007-19 Jan Riehme, Andrea Walther, Jörg Stiller, Uwe Naumann: Adjoints for Time-Dependent Optimal Control
2007-20 Joost-Pieter Katoen, Daniel Klink, Martin Leucker, and Verena Wolf: Three-Valued Abstraction for Probabilistic Systems
2007-21 Tingting Han, Joost-Pieter Katoen, and Alexandru Mereacre: Compositional Modeling and Minimization of Time-Inhomogeneous Markov Chains
2007-22 Heiner Ackermann, Paul W. Goldberg, Vahab S. Mirrokni, Heiko Röglin, and Berthold Vöcking: Uncoordinated Two-Sided Markets
2008-01 * Fachgruppe Informatik: Jahresbericht 2007/2008
2008-02 Henrik Bohnenkamp, Marielle Stoelinga: Quantitative Testing
2008-03 Carsten Fuhs, Jürgen Giesl, Aart Middeldorp, Peter Schneider-Kamp, René Thiemann, Harald Zankl: Maximal Termination
2008-04 Uwe Naumann, Jan Riehme: Sensitivity Analysis in Sisyphe with the AD-Enabled NAGWare Fortran Compiler
2008-05 Frank G. Radmacher: An Automata Theoretic Approach to the Theory of Rational Tree Relations
2008-06 Uwe Naumann, Laurent Hascoet, Chris Hill, Paul Hovland, Jan Riehme, Jean Utke: A Framework for Proving Correctness of Adjoint Message Passing Programs
2008-07 Alexander Nyßen, Horst Lichter: The MeDUSA Reference Manual, Second Edition
2008-08 George B. Mertzios, Stavros D. Nikolopoulos: The $\lambda$-cluster Problem on Parameterized Interval Graphs
2008-09 George B. Mertzios, Walter Unger: An optimal algorithm for the k-fixed-endpoint path cover on proper interval graphs
2008-10 George B. Mertzios, Walter Unger: Preemptive Scheduling of Equal-Length Jobs in Polynomial Time
2008-11 George B. Mertzios: Fast Convergence of Routing Games with Splittable Flows
2008-12 Joost-Pieter Katoen, Daniel Klink, Martin Leucker, Verena Wolf: Abstraction for stochastic systems by Erlang's method of stages
2008-13 Beatriz Alarcón, Fabian Emmes, Carsten Fuhs, Jürgen Giesl, Raúl Gutiérrez, Salvador Lucas, Peter Schneider-Kamp, René Thiemann: Improving Context-Sensitive Dependency Pairs
2008-14 Bastian Schlich: Model Checking of Software for Microcontrollers
2008-15 Joachim Kneis, Alexander Langer, Peter Rossmanith: A New Algorithm for Finding Trees with Many Leaves

2008-16 Hendrik vom Lehn, Elias Weingärtner and Klaus Wehrle: Comparing recent network simulators: A performance evaluation study
2008-17 Peter Schneider-Kamp: Static Termination Analysis for Prolog using Term Rewriting and SAT Solving
2008-18 Falk Salewski: Empirical Evaluations of Safety-Critical Embedded Systems
2008-19 Dirk Wilking: Empirical Studies for the Application of Agile Methods to Embedded Systems
2009-01 * Fachgruppe Informatik: Jahresbericht 2009
2009-02 Taolue Chen, Tingting Han, Joost-Pieter Katoen, Alexandru Mereacre: Quantitative Model Checking of Continuous-Time Markov Chains Against Timed Automata Specifications
2009-03 Alexander Nyßen: Model-Based Construction of Embedded Real-Time Software - A Methodology for Small Devices
2009-05 George B. Mertzios, Ignasi Sau, Shmuel Zaks: A New Intersection Model and Improved Algorithms for Tolerance Graphs
2009-06 George B. Mertzios, Ignasi Sau, Shmuel Zaks: The Recognition of Tolerance and Bounded Tolerance Graphs is NP-complete
2009-07 Joachim Kneis, Alexander Langer, Peter Rossmanith: Derandomizing Non-uniform Color-Coding I
2009-08 Joachim Kneis, Alexander Langer: Satellites and Mirrors for Solving Independent Set on Sparse Graphs
2009-09 Michael Nett: Implementation of an Automated Proof for an Algorithm Solving the Maximum Independent Set Problem
2009-10 Felix Reidl, Fernando Sánchez Villaamil: Automatic Verification of the Correctness of the Upper Bound of a Maximum Independent Set Algorithm
2009-11 Kyriaki Ioannidou, George B. Mertzios, Stavros D. Nikolopoulos: The Longest Path Problem is Polynomial on Interval Graphs
2009-12 Martin Neuhäußer, Lijun Zhang: Time-Bounded Reachability in Continuous-Time Markov Decision Processes
2009-13 Martin Zimmermann: Time-optimal Winning Strategies for Poset Games
2009-14 Ralf Huuck, Gerwin Klein, Bastian Schlich (eds.): Doctoral Symposium on Systems Software Verification (DS SSV'09)
2009-15 Joost-Pieter Katoen, Daniel Klink, Martin Neuhäußer: Compositional Abstraction for Stochastic Systems
2009-16 George B. Mertzios, Derek G. Corneil: Vertex Splitting and the Recognition of Trapezoid Graphs
2009-17 Carsten Kern: Learning Communicating and Nondeterministic Automata
2009-18 Paul Hänsch, Michaela Slaats, Wolfgang Thomas: Parametrized Regular Infinite Games and Higher-Order Pushdown Strategies
2010-01 * Fachgruppe Informatik: Jahresbericht 2010
2010-02 Daniel Neider, Christof Löding: Learning Visibly One-Counter Automata in Polynomial Time
2010-03 Holger Krahn: MontiCore: Agile Entwicklung von domänenspezifischen Sprachen im Software-Engineering

| | |
|---|---|
| 2010-04 | René Wörzberger: Management dynamischer Geschäftsprozesse auf Basis statischer Prozessmanagementsysteme |
| 2010-05 | Daniel Retkowitz: Softwareunterstützung für adaptive eHome-Systeme |
| 2010-06 | Taolue Chen, Tingting Han, Joost-Pieter Katoen, Alexandru Mereacre: Computing maximum reachability probabilities in Markovian timed automata |
| 2010-07 | George B. Mertzios: A New Intersection Model for Multitolerance Graphs, Hierarchy, and Efficient Algorithms |
| 2010-08 | Carsten Otto, Marc Brockschmidt, Christian von Essen, Jürgen Giesl: Automated Termination Analysis of Java Bytecode by Term Rewriting |
| 2010-09 | George B. Mertzios, Shmuel Zaks: The Structure of the Intersection of Tolerance and Cocomparability Graphs |
| 2010-10 | Peter Schneider-Kamp, Jürgen Giesl, Thomas Ströder, Alexander Serebrenik, René Thiemann: Automated Termination Analysis for Logic Programs with Cut |
| 2010-11 | Martin Zimmermann: Parametric LTL Games |
| 2010-12 | Thomas Ströder, Peter Schneider-Kamp, Jürgen Giesl: Dependency Triples for Improving Termination Analysis of Logic Programs with Cut |
| 2010-13 | Ashraf Armoush: Design Patterns for Safety-Critical Embedded Systems |
| 2010-14 | Michael Codish, Carsten Fuhs, Jürgen Giesl, Peter Schneider-Kamp: Lazy Abstraction for Size-Change Termination |
| 2010-15 | Marc Brockschmidt, Carsten Otto, Christian von Essen, Jürgen Giesl: Termination Graphs for Java Bytecode |
| 2010-16 | Christian Berger: Automating Acceptance Tests for Sensor- and Actuator-based Systems on the Example of Autonomous Vehicles |
| 2010-17 | Hans Grönniger: Systemmodell-basierte Definition objektbasierter Modellierungssprachen mit semantischen Variationspunkten |
| 2010-18 | Ibrahim Armaç: Personalisierte eHomes: Mobilität, Privatsphäre und Sicherheit |
| 2010-19 | Felix Reidl: Experimental Evaluation of an Independent Set Algorithm |
| 2010-20 | Wladimir Fridman, Christof Löding, Martin Zimmermann: Degrees of Lookahead in Context-free Infinite Games |
| 2011-01 * | Fachgruppe Informatik: Jahresbericht 2011 |
| 2011-02 | Marc Brockschmidt, Carsten Otto, Jürgen Giesl: Modular Termination Proofs of Recursive Java Bytecode Programs by Term Rewriting |
| 2011-03 | Lars Noschinski, Fabian Emmes, Jürgen Giesl: A Dependency Pair Framework for Innermost Complexity Analysis of Term Rewrite Systems |
| 2011-04 | Christina Jansen, Jonathan Heinen, Joost-Pieter Katoen, Thomas Noll: A Local Greibach Normal Form for Hyperedge Replacement Grammars |
| 2011-07 | Shahar Maoz, Jan Oliver Ringert, Bernhard Rumpe: An Operational Semantics for Activity Diagrams using SMV |
| 2011-08 | Thomas Ströder, Fabian Emmes, Peter Schneider-Kamp, Jürgen Giesl, Carsten Fuhs: A Linear Operational Semantics for Termination and Complexity Analysis of ISO Prolog |
| 2011-09 | Markus Beckers, Johannes Lotz, Viktor Mosenkis, Uwe Naumann (Editors): Fifth SIAM Workshop on Combinatorial Scientific Computing |
| 2011-10 | Markus Beckers, Viktor Mosenkis, Michael Maier, Uwe Naumann: Adjoint Subgradient Calculation for McCormick Relaxations |
| 2011-11 | Nils Jansen, Erika Ábrahám, Jens Katelaan, Ralf Wimmer, Joost-Pieter Katoen, Bernd Becker: Hierarchical Counterexamples for Discrete-Time Markov Chains |
| 2011-12 | Ingo Felscher, Wolfgang Thomas: On Compositional Failure Detection in Structured Transition Systems |
| 2011-13 | Michael Förster, Uwe Naumann, Jean Utke: Toward Adjoint OpenMP |
| 2011-14 | Daniel Neider, Roman Rabinovich, Martin Zimmermann: Solving Muller Games via Safety Games |
| 2011-16 | Niloofar Safiran, Uwe Naumann: Toward Adjoint OpenFOAM |
| 2011-18 | Kamal Barakat: Introducing Timers to pi-Calculus |
| 2011-19 | Marc Brockschmidt, Thomas Ströder, Carsten Otto, Jürgen Giesl: Automated Detection of Non-Termination and NullPointerExceptions for Java Bytecode |
| 2011-24 | Callum Corbett, Uwe Naumann, Alexander Mitsos: Demonstration of a Branch-and-Bound Algorithm for Global Optimization using McCormick Relaxations |
| 2011-25 | Callum Corbett, Michael Maier, Markus Beckers, Uwe Naumann, Amin Ghobeity, Alexander Mitsos: Compiler-Generated Subgradient Code for McCormick Relaxations |
| 2011-26 | Hongfei Fu: The Complexity of Deciding a Behavioural Pseudometric on Probabilistic Automata |
| 2012-01 * | Fachgruppe Informatik: Annual Report 2012 |
| 2012-02 | Thomas Heer: Controlling Development Processes |

* These reports are only available as a printed version.

Please contact biblio@informatik.rwth-aachen.de to obtain copies.